\documentclass[a4paper,twocolumn,11pt,accepted=2020-12-21]{quantumarticle}
\pdfoutput=1
\usepackage[utf8]{inputenc}
\usepackage[english]{babel}
\usepackage[T1]{fontenc}

\usepackage[numbers,sort&compress]{natbib}

\usepackage{graphicx}
\usepackage{amsmath}
\usepackage{amsfonts}
\usepackage{amssymb}
\usepackage{amsthm}

\usepackage{mathtools}

\usepackage{stmaryrd}

\usepackage{dsfont}
\usepackage{microtype}

\usepackage[breaklinks=true,colorlinks]{hyperref}

\newcommand{\cf}{cf.\ }

\newcommand{\bd}{\begin{equation*}}
\newcommand{\ed}{\end{equation*}}

\newcommand{\Tr}{\operatorname{Tr}}

\newcommand{\indexc}{\mathrm{c}}
\newcommand{\indexh}{\mathrm{h}}

\newcommand{\betac}{\beta_\indexc}

\newcommand{\Tc}{T_\indexc}
\newcommand{\Th}{T_\indexh}

\newcommand{\sta}{\mathrm{STA}}
\newcommand{\cd}{\mathrm{CD}}
\newcommand{\ad}{\mathrm{ad}}

\newcommand{\Hcd}{H_\cd}

\newcommand{\Qc}{Q_\indexc}
\newcommand{\Qh}{Q_\indexh}
\newcommand{\lambdac}{\lambda_\indexc}
\newcommand{\lambdah}{\lambda_\indexh}

\begin{document}

\title{Multi-spin counter-diabatic driving in many-body quantum Otto refrigerators}

\author{Andreas Hartmann}
\email{andreas.Hartmann@uibk.ac.at}
\affiliation{Institut f\"ur Theoretische Physik, Universit\"at Innsbruck, Technikerstra{\ss}e~21a, A-6020~Innsbruck, Austria}
\author{Victor Mukherjee}
%\email{mukherjeev@iiserbpr.ac.in}
\affiliation{Department of Physical Sciences, IISER Berhampur, Berhampur 760010, India}
\author{Glen Bigan Mbeng}
\email{glen.mbeng@uibk.ac.at}
\affiliation{Institut f\"ur Theoretische Physik, Universit\"at Innsbruck, Technikerstra{\ss}e~21a, A-6020~Innsbruck, Austria}
\author{Wolfgang Niedenzu}
\author{Wolfgang Lechner}
%\email{Wolfgang.Lechner@uibk.ac.at}
\affiliation{Institut f\"ur Theoretische Physik, Universit\"at Innsbruck, Technikerstra{\ss}e~21a, A-6020~Innsbruck, Austria}

\begin{abstract}
Quantum refrigerators pump heat from a cold to a hot reservoir. In the few-particle regime, counter-diabatic (CD) driving of, originally adiabatic, work-exchange strokes is a promising candidate to overcome the bottleneck of vanishing cooling power. Here, we present a finite-time many-body quantum refrigerator that yields finite cooling power at high coefficient of performance, that considerably outperforms its non-adiabatic counterpart. 
We employ multi-spin CD driving and numerically investigate the scaling behavior of the refrigeration performance with system size.
We further prove that optimal refrigeration via the exact CD protocol is a catalytic process.
\end{abstract}

\date{December 21, 2020}

\maketitle

\section{Introduction}
Heat engines and refrigerators are a cornerstone of modern physics and indispensable in today's society \cite{cengelbook}. Unravelling their fundamental laws in the few-particle regime has lead to the study of so-called quantum heat engines and refrigerators \cite{alicki1979quantum, kosloff1984quantum, kosloff2013quantum, gelbwaser2015thermodynamics, vinjanampathy2016quantum, karimi2016otto, kosloff2017quantum, binder2019thermodynamicsbook}. With the recent progress in controling quantum systems~\cite{bernien2017probing, choi2016exploring, bordia2017probing}, such quantum heat engines could already be experimentally realized using various single-body quantum working media (WM)~\cite{koski2014experimental, rossnagel2016single, klaers2017squeezed, peterson2019experimental, vonlindenfels2019spin, klatzow2019experimental}. Whereas heat engines convert thermal energy into work, their counterparts, namely refrigerators, cool down a cold bath by pumping heat from the cold to the hot reservoir, thereby consuming work~\cite{rezek2009the, levy2012quantum, yuan2014coefficient, long2015performance, abah2016optimal, karimi2016otto, kosloff2017quantum, niedenzu2019quantized}. The maximum coefficient of performance (CoP) of refrigerators is limited by the Carnot CoP~\cite{callen1985thermodynamics}. For these infinitely long (adiabatic) cycle times, the corresponding cooling power naturally converges to zero. A natural question thus arises whether such quantum refrigerators can be driven in finite time, yet produce a finite cooling power.

\par 

So called shortcuts to adiabaticity (STA)~\cite{arimondo2013chapter, delcampo2015controlling, delcampo2019focus, guery-odelin2019shortcuts} are a promising candidate to overcome this fundamental bottleneck, due to minimizing quantum friction~\cite{kosloff2002discrete, feldmann2003quantum, feldmann2006quantum} during the work-exchange strokes. These STA methods~\cite{demirplak2003adiabatic, berry2009transitionless, chen2010fast, chen2011lewis, takahashi2013transitionless, jarzynski2013generating} including counter-diabatic (CD) driving~\cite{delcampo2013shortcuts, jarzynski2013generating, damski2014counterdiabatic, sels2017minimizing, claeys2019floquet, hartmann2019rapid} where an additional CD Hamiltonian is applied to suppress any transitions between the system's eigenstates during the Hamiltonian's dynamics, have recently been applied in the field of quantum thermodynamics to enhance the performance of quantum heat engines~\cite{delcampo2014more, abah2018performance, abah2019shortcut, dupays2020superadiabatic, cakmak2019spin}  and refrigerators~\cite{funo2019speeding, abah2020shortcut} using \emph{single-body} quantum WM. For the latter, \emph{exact} local CD terms can be found analytically~\cite{chen2010fast, takahashi2013transitionless}.

\par 

In general, identifying the \emph{exact} CD term requires \emph{a priori knowledge of the system's eigenstates} at all times during the Hamiltonian's dynamics and which is numerically and experimentally impracticable for \emph{many-body} WM. With this challenge in mind, Sels and Polkovnikov \cite{sels2017minimizing} have developed a variational principle where \emph{approximate} multi-spin CD terms can be found.
Based on this method, a quantum heat engine using a many-body quantum WM and local 1-spin CD Hamiltonian could be implemented~\cite{hartmann2020manybody}. However, the latter are poor approximations and can only efficiently speed up an Otto cycle with weakly coupled WM. The question whether a cycle with WM with complex all-to-all connectivity -- that provides versatile applications in state-of-the-art experimental setups in the field of quantum computation and simulation, \cf Refs.~\cite{cirac2012goals, georgescu2014quantum, LucasAnnealing, albash2018adiabatic} -- can be sped-up is thus still open.

\par 

In this work, we present a \emph{finite-time} many-body quantum refrigerator (QR) with finite cooling power using additional approximate \emph{multi-spin} CD terms. The WM is described by an all-to-all connected Ising spin system and can efficiently be driven using non-local, yet well approximated $p$-spin CD terms. We numerically demonstrate a large enhancement in cooling power and CoP along with improved scaling behaviour of the cooling power with the system size for the sped-up QR compared to its original non-adiabatic and traditional $1$-spin CD counterparts. For the QR with single-body quantum WM, we find an analytical expression for the CoP. For the many-body WM, we provide an analytical proof that \emph{exact} CD driving implies a zero work component of the additional external control device and thus is a catalytic process. Remarkably, the latter fully assists the piston to run the QR in this case, mimicking the adiabatic quantum cycle, yet in finite time. 

\par 

This work is structured as follows: In Sec.~\ref{sec_m}, we introduce the quantum Otto refrigerator using a many-body spin system as its quantum WM and present multi-spin local CD driving. In Sec.~\ref{sec_res}, we numerically analyze the performance of the corresponding refrigerators and conclude our results in Sec.~\ref{sec_dis} while giving an outlook on future research.

\section{Methods}\label{sec_m}
\subsection{Quantum Otto refrigerator}\label{subsec_qr}
Quantum Otto refrigerators cyclically pump heat from a cold to a hot reservoir by consuming work. Its corresponding four-stroke quantum Otto cycle~\cite{kosloff2017quantum} consists of two heat-exchange strokes  -- where the quantum WM is alternatingly coupled to two heat baths -- and two work-exchange strokes. 

\begin{figure}
  \centering
  \includegraphics[width=.95\columnwidth]{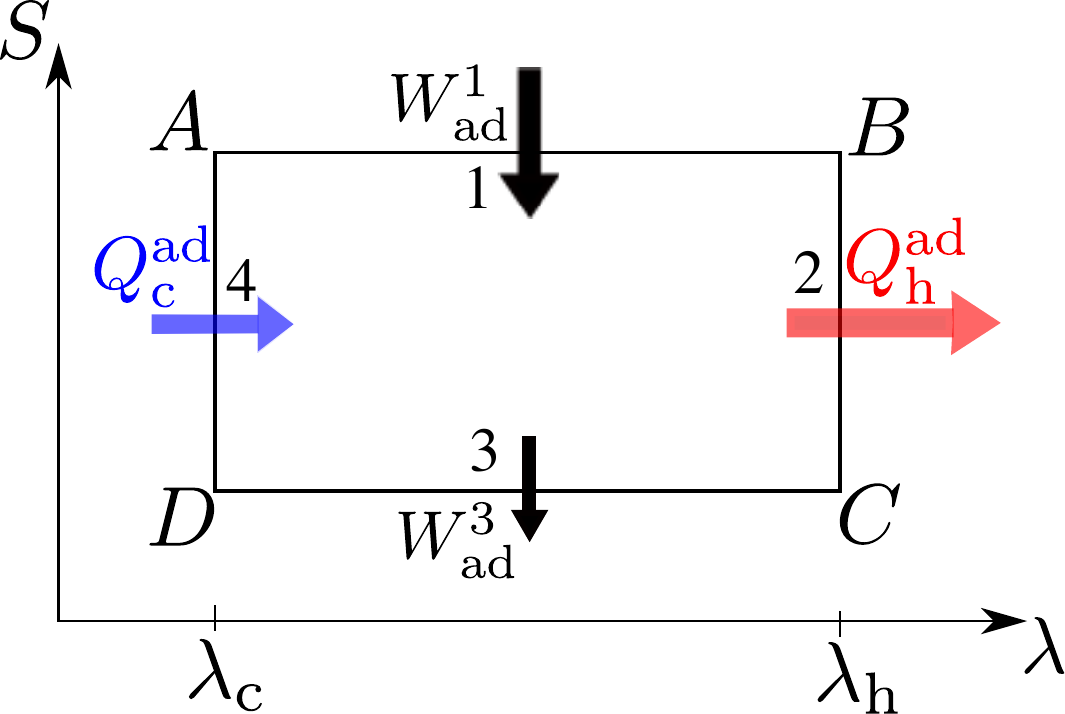}
  \caption{\textbf{Entropy-working parameter diagram of quantum Otto refrigerator.} The quantum Otto cycle consists of two adiabatic (1 and 3) and two thermal strokes (2 and 4). At the end of the cycle, the heat $\Qc^\mathrm{ad}$ is pumped from the cold bath at temperature $\Tc$ into the hot bath at temperature $\Th$ by consuming the work $W^1_\mathrm{ad}+W^3_\mathrm{ad} > 0$.}
  \label{fig:fig1}
\end{figure}

The four strokes are (\cf Fig.~\ref{fig:fig1}):
\begin{enumerate}
\item \emph{Adiabatic compression} ($A \to B$): Initially, at cold temperature $\Tc=1/\betac$, Hamiltonian $H_0(\lambdac)$ with working parameter $\lambdac \coloneqq \lambda(t=0)$ and in the thermal state $\rho_A=e^{-H_0(\lambdac)/\Tc}/ \Tr[e^{-H_0(\lambdac)/\Tc}]$, the WM is changed adiabatically during the first isentropic stroke of duration $\tau_1$ with final Hamiltonian $H_0(\lambdah)$ and mixed state $\rho_B$ with $\lambdah \coloneqq \lambda(t=\tau)$.

\item \emph{Hot isochore} ($B \to C$): The WM is brought in contact with the hot thermal bath at temperature $\Th$ until it equilibrates to the thermal state $\rho_C=e^{-H_0(\lambdah)/\Th}/ \Tr[e^{-H_0(\lambdah)/\Th}]$. During this heat-exchange stroke of duration $\tau_2$, the heat $\Qh$ is imparted to the hot bath and the WM's Hamiltonian $H_0(\lambdah)$ remains unchanged.

\item \emph{Adiabatic expansion} ($C \to D$): The Hamiltonian $H_0(\lambdah)$ changes back to $H_0(\lambdac)$ during the stroke duration $\tau_3$ until the WM attains the state $\rho_D$.

\item \emph{Cold isochore} ($D \to A$): The WM is brought in contact with the cold bath at temperature $\Tc$ and working parameter $\lambdac$ and cools down during the duration $\tau_4$ to its originally initial thermal state $\rho_A$.
\end{enumerate}

During one cycle, the WM extracts the heat $\Qc^\mathrm{ad} \coloneqq \langle H_0(\lambdac) \rangle_{\rho_A} - \langle H_0(\lambdac) \rangle_{\rho_D} > 0$ from the cold thermal bath, thereby consumes the work $W^{1+3}_\mathrm{ad} \coloneqq W^1_\mathrm{ad}  + W^3_\mathrm{ad}  > 0$ (here, $W^1_\mathrm{ad} = \langle H_0(\lambdah) \rangle_{\rho_B} - \langle H_0(\lambdac) \rangle_{\rho_A}$ and $W^3_\mathrm{ad} = \langle H_0(\lambdac) \rangle_{\rho_D} - \langle H_0(\lambdah) \rangle_{\rho_C}$). A positive (negative) sign of the work components corresponds to work performed on (extracted from) the WM. Analogously, a positive (negative) sign of the heat corresponds to heat extracted from (imparted to) the thermal bath.

The cooling power is defined as the pumped heat $\Qc^\mathrm{ad}$ over the total cycle time $\tau_{\mathrm{cycle}}=\sum_{l=1}^4 \tau_l$, i.e.,
\begin{equation}
  \mathcal{J}_\ad=\dfrac{\mathrm{pumped \,heat}}{\mathrm{cycle \,duration}}=\dfrac{\Qc^\mathrm{ad}}{\tau_{\mathrm{cycle}}}.
  \label{eq_power_ad}
\end{equation}

The CoP of the Otto cycle is defined as the heat $\Qc^\mathrm{ad}$ pumped from the cold bath over the consumed work $W_\mathrm{ad}^1+W_\mathrm{ad}^3$, i.e.,
\begin{equation}
\epsilon_\mathrm{ad}=\dfrac{\mathrm{pumped \,heat}}{\mathrm{consumed \,work}}= \dfrac{\Qc^\mathrm{ad}}{W_\mathrm{ad}^1+W_\mathrm{ad}^3}.
\label{eq_Cop_ad}
\end{equation}

In the adiabatic limit -- where the isentropic strokes with durations $\tau_1$ and $\tau_3$ are infinitely long -- the cooling power of these \emph{adiabatic} quantum refrigerators goes to zero, i.e., $\underset{\tau_1,\tau_3 \to \infty}{\lim} \mathcal{J}_\ad \to 0$.
To overcome this bottleneck, one can implement so called STA techniques for the two work-exchange strokes.

\subsection{Quantum working medium}\label{subsec_wm}
As our quantum WM, we consider an all-to-all connected Ising spin chain with Hamiltonian
\begin{align}
H_0(t) &= - [1 - \vartheta(t)] \sum_{j=1}^N h_j \sigma_j^x \nonumber \\
&- \vartheta(t) \left[\sum_{j=1}^N b_j \sigma_j^z + \sum_{j=1}^N \sum_{k < j} J_{jk} \sigma_j^z \sigma_k^z \right]
\label{eq_H0}
\end{align}
where $h_j$ and $b_j$, respectively, are the time-dependent transverse and longitudinal magnetic field strengths at site $j$ and $J_{jk}$ the interaction strength between spins at sites $j$ and $k$. $\vartheta(t)$ is a continuous function that fulfills $\vartheta(t=0)=0$ and $\vartheta(t=\tau_1)=1$. For the second isentropic stroke the initial and final values of the function $\vartheta(t)$ are interchanged. 

Throughout this work, we parametrize the working parameters $\lambdac$ and $\lambdah$ with the magnetic fields and interaction strengths at each point of Fig.~\ref{fig:fig1}, i.e. $\lambdac \coloneqq \{h_{j, \mathrm{i}}, b_{j, \mathrm{i}}, J_{jk, \mathrm{i}} \}$ and $\lambdah \coloneqq \{h_{j, \mathrm{f}}, b_{j, \mathrm{f}}, J_{jk, \mathrm{f}} \}$. Here, $h_{j, \mathrm{i}}$, $b_{j, \mathrm{i}}$ and $J_{jk, \mathrm{i}}$ are the values of the magnetic fields and interaction strengths at points $A$ and $D$, and $h_{j, \mathrm{f}}$, $b_{j, \mathrm{f}}$ and $J_{jk, \mathrm{f}}$ at points $B$ and $C$, respectively. The explicit forms of the functions $h_j(t)$, $b_j(t)$ and $J_{jk}(t)$ as well as the sweep function $\vartheta(t)$ are given in Appendix~\ref{sec:A}.

\subsection{Multi-spin counter-diabatic driving}\label{subsec_cd}
The underlying idea of CD driving~\cite{delcampo2013shortcuts, jarzynski2013generating, damski2014counterdiabatic, sels2017minimizing, claeys2019floquet, hartmann2019rapid} is to efficiently drive an adiabatic evolution of a Hamiltonian in \emph{finite time} by suppressing transitions between its eigenstates. Thus, we always track the instantaneous eigenstates during the whole sweep.
Finding the \emph{exact} CD term requires a priori knowledge of the system's eigenstates for every time during the sweep which is numerically and experimentally challenging.

\par

In order to overcome this bottleneck, an analytical variational principle has been developed recently~\cite{sels2017minimizing, claeys2019floquet} to find \emph{approximate} CD terms.\\ 
In this work, we drive the WM during the isentropic strokes with the total Hamiltonian
\begin{equation}
H_{\mathrm{STA}}(t)=H_0(t)+H_{\mathrm{CD}}(t)
\label{eq_Hsta}
\end{equation}
where $H_0(t)$, Eq.~\eqref{eq_H0}, is the original non-adiabatic and $\Hcd(t)$ the additional CD Hamiltonian that suppresses coherences in the WM that cause quantum friction in finite-time sweeps~\cite{kosloff2002discrete, feldmann2003quantum, feldmann2006quantum}. 

The additional CD Hamiltonian reads
\begin{equation}
H_\mathrm{CD}(t) = \dot{\vartheta}(t) \mathcal{A}_\vartheta(t)
\label{eq_Hcd}
\end{equation}
with $\mathcal{A}_\vartheta(t)$ the \emph{exact} adiabatic gauge potential~\cite{sels2017minimizing, kolodrubetz2017geometry, claeys2019floquet} and $\dot{\vartheta}(t)$ the derivative of the sweep function of Eq.~\eqref{eq_H0}. We rely on an \emph{approximate} adiabatic gauge potential $\mathcal{A}^*_\vartheta$ that contains $p$-spin terms (with $p \leq N$) and an odd number of $\sigma^y$ terms (e.g., $\sigma_j^y$ for $p=1$, $\sigma_j^y$, $\sigma_j^y \sigma_k^x$ and $\sigma_j^y \sigma_k^z$ for $p=2$, etc.). For $p=N$, we obtain the solution of the \emph{exact} adiabatic gauge potential that entails all combinations of $N$-spin terms (\cf Ref.~\cite{delcampo2012assisted} in the case of quantum criticality). 
As an example, for $p=1$ we apply the ansatz $\mathcal{A}^*_\vartheta = \sum_{j=1}^N \alpha_j \sigma_j^y$ and solve for the optimal solution of each coefficient $\alpha_j$ by minimizing the operator distance between the exact and approximate adiabatic gauge potential.
For more details, see Appendix~\ref{sec:B} and Refs.~\cite{sels2017minimizing, kolodrubetz2017geometry, claeys2019floquet}. 
\par 
We note, that we apply the CD Hamiltonian \emph{only} in the isentropic strokes as these are normally much longer than the thermalization strokes. However, techniques to speed up the latter have been also developed recently~\cite{Alipour2020shortcutsto, dupays2020superadiabatic, dann2020fast, dann2019shortcut, das2020quantum}.

\subsection{Quantum refrigerator under STA}\label{subsec_wc}
The introduction of the additional \emph{approximate} CD term $H^*_\mathrm{CD}(t)$ in the Hamiltonian's dynamics during the, originally non-adiabatic, work-exchange strokes requires a careful definition of work, cooling power and CoP.
In Ref.~\cite{hartmann2020manybody}, the division
\begin{equation}
W^l_\mathrm{STA} = W^l_\mathrm{0} + W^l_\mathrm{CD}
\end{equation}
with $W^l_\mathrm{STA} \equiv \Delta E = \int_0^{\tau_l} \mathrm{Tr} \left[\rho(t) \dot{H}_\mathrm{STA}(t)\right] dt$ the total exchanged work for each of the two work-exchange strokes $l \in \{1,3\}$ with duration $\tau_l$ has been introduced.
The corresponding work components thus read
\begin{align}
W_0^l &= \int_0^{\tau_l} \mathrm{Tr} \left[\rho(t) \dot{H}_0(t)\right] dt\label{eq_W0}, \\
W_\mathrm{CD}^l &= \int_0^{\tau_l} \mathrm{Tr} \left[\rho(t) \dot{H}_\mathrm{CD}(t)\right] dt
\label{eq_Wcd}
\end{align}
where the work $W_0^l$ stems from the piston and $W_\mathrm{CD}^l$ from the external control device that implements $H^*_\mathrm{CD}(t)$. 

\par 

As we will show in more detail in Appendix~\ref{sec:C}, this work component $W^l_\mathrm{CD}$ is \emph{zero} if the additional adiabatic gauge potential and thus the CD Hamiltonian $H^*_\mathrm{CD}(t)$ is \emph{exact}, i.e., $\mathcal{A}^*_\vartheta(t) = \mathcal{A}_\vartheta(t)$ and thus $H^*_\mathrm{CD}(t)=H_\mathrm{CD}(t)$. 
Hence, the work component $W^{1+3}_\mathrm{CD}$ stemming from the external control device during one cycle can be seen as an artifact of \emph{inexact}, e.g. $1$-spin CD driving (\cf Ref.~\cite{hartmann2020manybody}). Experimentally it is advantageous to have a vanishing contribution from the external control device as we want the external control device to fully assist the piston instead of just exploiting it to run the quantum refrigerator. In other words, the \emph{exact} CD drive is \emph{catalytic} in the sense that it allows for speeding up the cycle without the external control device being altered (charged or discharged) after a cycle. Therefore, as we will see later, applying multi-spin CD terms can efficiently speed up the cycle without excessively exploiting the external control device.

\par 

The cooling power under STA is given by
\begin{equation}
  \mathcal{J}_\sta \coloneqq \dfrac{\mathrm{pumped \,heat}}{\mathrm{cycle \,duration}}=\dfrac{\Qc}{\tau_{\mathrm{cycle}}}
  \label{eq_power_sta}
\end{equation}
and the CoP reads
\begin{equation}
\epsilon_\mathrm{STA} \coloneqq \dfrac{\mathrm{pumped \,heat}}{\mathrm{consumed \,work}}= \dfrac{\Qc}{W^1_\mathrm{STA}+W^3_\mathrm{STA}}
\label{eq_cop_sta}
\end{equation}
where $W_\mathrm{STA}^{1+3} = W_\mathrm{STA}^{1} + W_\mathrm{STA}^{3} > 0$ is the total work performed on (consumed by) the WM per cycle. Note the difference between the pumped heat $\Qc^\mathrm{ad}$ for the adiabatic [\cf Eqs.~\eqref{eq_power_ad} and \eqref{eq_Cop_ad}] and $\Qc$ for the sped-up cycle [\cf Eqs.~\eqref{eq_power_sta} and \eqref{eq_cop_sta}].

For the case of a single-body quantum WM that is modelled by Eq.~\eqref{eq_H0} with $N=1$, i.e., this reduces to the Landau-Zener (LZ) model, the CoP evaluates to
\begin{equation}
\epsilon_\mathrm{LZ} \coloneqq \dfrac{h_\mathrm{x,i}}{b_\mathrm{z,f} - h_\mathrm{x,i}}
\label{eq_CoP_LZ}
\end{equation}
where $h_\mathrm{x,i}$ is the initial value of the transverse magnetic field in the first isentropic stroke and $b_\mathrm{z,f}$ the final value of the longitudinal magnetic field strength, respectively. This expression is positive provided that $b_\mathrm{z,f} > h_\mathrm{x,i}$ and is limited by the Carnot CoP $\epsilon_\mathrm{C}=\Tc/(\Th - \Tc)$~\cite{callen1985thermodynamics} (see Appendix~\ref{sec:D} for more details).

\section{Numerical Results}\label{sec_res}
In this section, we present the numerical results of the proposed quantum Otto refrigerator. To this end, we used the QuTip 4.2~\cite{johannson2013qutip} Python package to simulate the quantum Otto cycle with (i)~non-adiabatic [$H_0(t)$, Eq.~\eqref{eq_H0}] and (ii)~STA Hamiltonian [$H^*_\mathrm{STA}(t)$, Eq.~\eqref{eq_Hsta}] with $p$-spin CD terms [$H^*_\mathrm{CD}(t)$, Eq.~\eqref{eq_Hcd}]. We numerically solved the von~Neumann equation for the isentropic strokes and computed the heat $\Qc$ and $\Qh$ for the two thermalization strokes as the energy difference between points $B$ and $C$ as well as $D$ and $A$ (\cf Fig.~\ref{fig:fig1}), respectively.
Throughout this numerical performance analysis, the temperatures for the cold and hot bath are set to $\Tc = 0.2$ and $\Th = 0.4$, respectively. The values for the working parameters at points $A$ and $D$ as well as $B$ and $C$ in Fig.~\ref{fig:fig1} read $\lambdac = \{ h_{j, \mathrm{i}} = 0.2, \, b_{j, \mathrm{i}} = 0, J_{jk, \mathrm{i}} = 0 \}$ and $\lambdah = \{ h_{j, \mathrm{f}} = 0, \, b_{j, \mathrm{f}} = 0.5, J_{jk, \mathrm{f}}= 0.1 \}$, respectively, for both, non-adiabatic $H_0(t)$ and STA Hamiltonian $H^*_\mathrm{STA}(t)$ with $p$-spin CD Hamiltonian $H^*_\mathrm{CD}(t)$ for different $p$. 
At each point $A$, $B$, $C$ and $D$, the additionally applied CD Hamiltonian is zero, i.e., $H^*_\mathrm{CD}(t=\sum_j \tau_j)=0$ where $j \in \{0, 1, 2, 3 \}$ and $\tau_0=0$.  The durations $\tau_2$ and $\tau_4$ of each thermalization stroke are set to $0.1$.

\subsection{Scaling with system size}
Our primary goal is to speed up the quantum refrigerator, i.e., we want to pump as much heat $\Qc$ as possible from the cold to the hot reservoir in the shortest amount of time. Thus, we are particularly interested in the cooling power $\mathcal{J}$ during one cycle and its scaling behaviour for different system sizes $N$.

\begin{figure}
  \centering
  \includegraphics[width=.95\columnwidth]{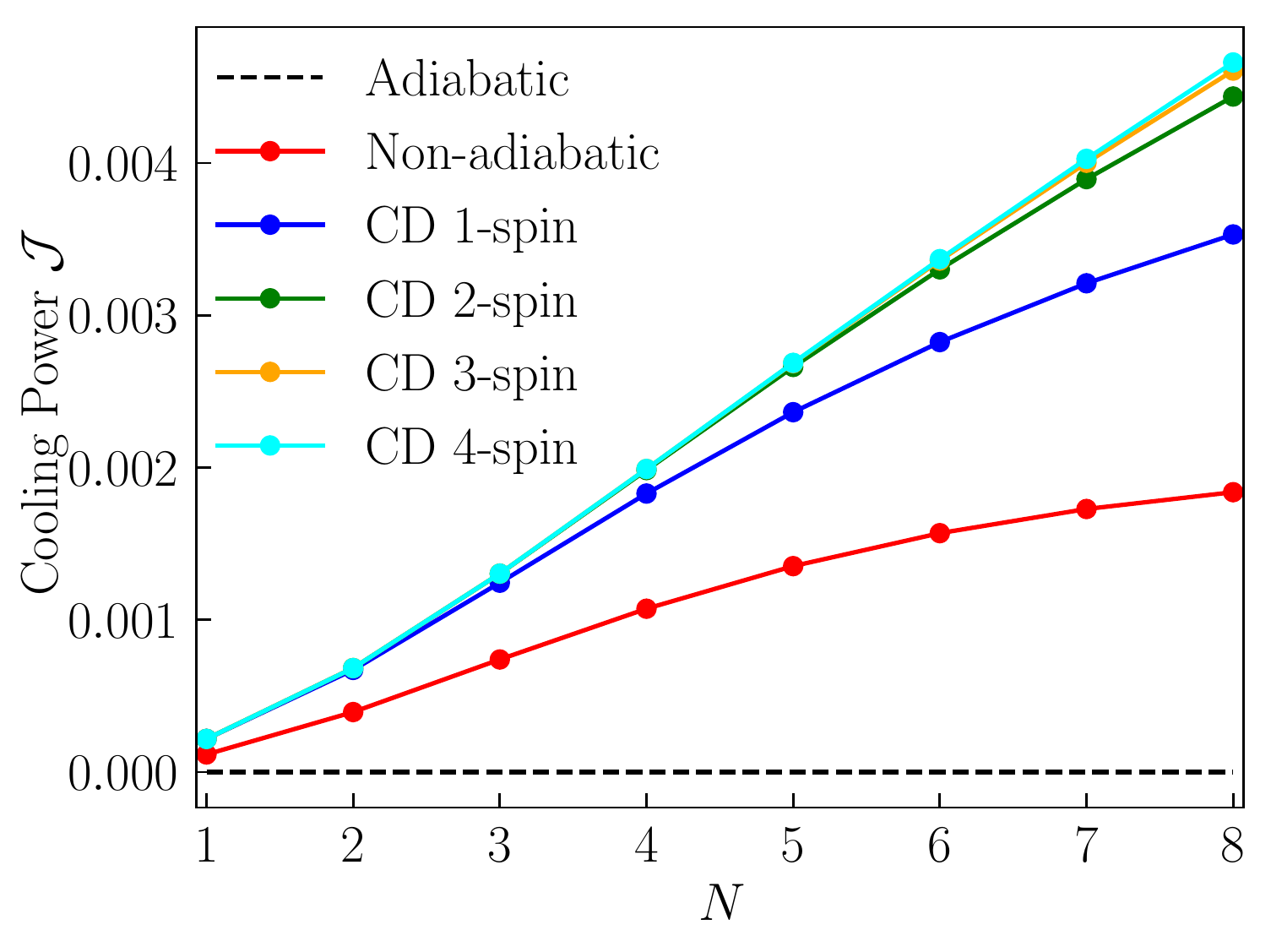}
  \caption{\textbf{Cooling power $\pmb{\mathcal{J}}$ over system size $\pmb{N}$.} Cooling power $\mathcal{J}$ of a quantum Otto cycle for isentropic stroke duration $\tau=\tau_1=\tau_3 \approx 40$ for (i)~non-adiabatic Hamiltonian [$H_0(t)$, Eq.~\eqref{eq_H0}] (red, bottom line) and (ii)~STA Hamiltonian [$H^*_\mathrm{STA}(t)$, Eq.~\eqref{eq_Hsta}] for different $p$-spin CD terms [$H^*_\mathrm{CD}(t)$, Eq.~\eqref{eq_Hcd}] over different system sizes $N$. Parameters: Cold and hot bath at temperatures $\Tc = 0.2$ and $\Th = 0.4$, respectively, and working parameters $\lambdac$ with $h_{j, \mathrm{i}}=0.2$, $b_{j, \mathrm{i}}=0$ and $J_{jk, \mathrm{i}}=0$ and $\lambdah$ with $h_{j, \mathrm{f}}=0$, $b_{j, \mathrm{f}}=0.5$ and $J_{jk, \mathrm{f}}=0.1$, magnetic field and interaction strengths, respectively. Duration of the thermalization strokes: $\tau_2=\tau_4=0.1$. Black-dashed line denotes cooling power at adiabatic limit where $\tau=\tau_1=\tau_3 \to \infty$.}
  \label{fig:fig2}
\end{figure}

Figure~\ref{fig:fig2} depicts the cooling power $\mathcal{J}$ for (i)~non-adiabatic [$H_0(t)$, Eq.~\eqref{eq_H0}] and (ii)~STA Hamiltonian [$H^*_\mathrm{STA}$, Eq.~\eqref{eq_Hsta}] for different $p$-spin CD terms [$H^*_\mathrm{CD}$, Eq.~\eqref{eq_Hcd}] over different system sizes $N$ of spins in the WM for an isentropic stroke duration of $\tau=\tau_1=\tau_3 \approx 40$. By applying $p$-spin CD terms, we can efficiently enhance the cooling power of our sped-up refrigerator by increasing the system size $N$ compared to the non-adiabatic counterpart. The relative enhancement in cooling power decreases the higher $p$ becomes. The major relative enhancement can be made by applying 1-spin or 2-spin CD terms. Including multi-spin terms (e.g. $3$ and $4$-body in the cases studied) only give a relatively slight improvement compared to 1-spin or 2-spin CD driving. We note, that there is a trade-off between enhanced cooling power with increasing $p$ and implementation complexity for an experiment. In the adiabatic limit (black-dashed line), the cooling power naturally converges to zero. 

From a practical point of view, we deem it more favorable if the increased cooling power $\mathcal{J}_\sta$ is due to the piston rather than exploiting the external control device. Consequently, we are interested in the work component $W^{1+3}_\mathrm{CD}$, Eq.~\eqref{eq_Wcd}. As shown in Appendix~\ref{sec:C} the latter is zero if the applied CD Hamiltonian is \emph{exact}. 

\begin{figure}
  \centering
  \includegraphics[width=.95\columnwidth]{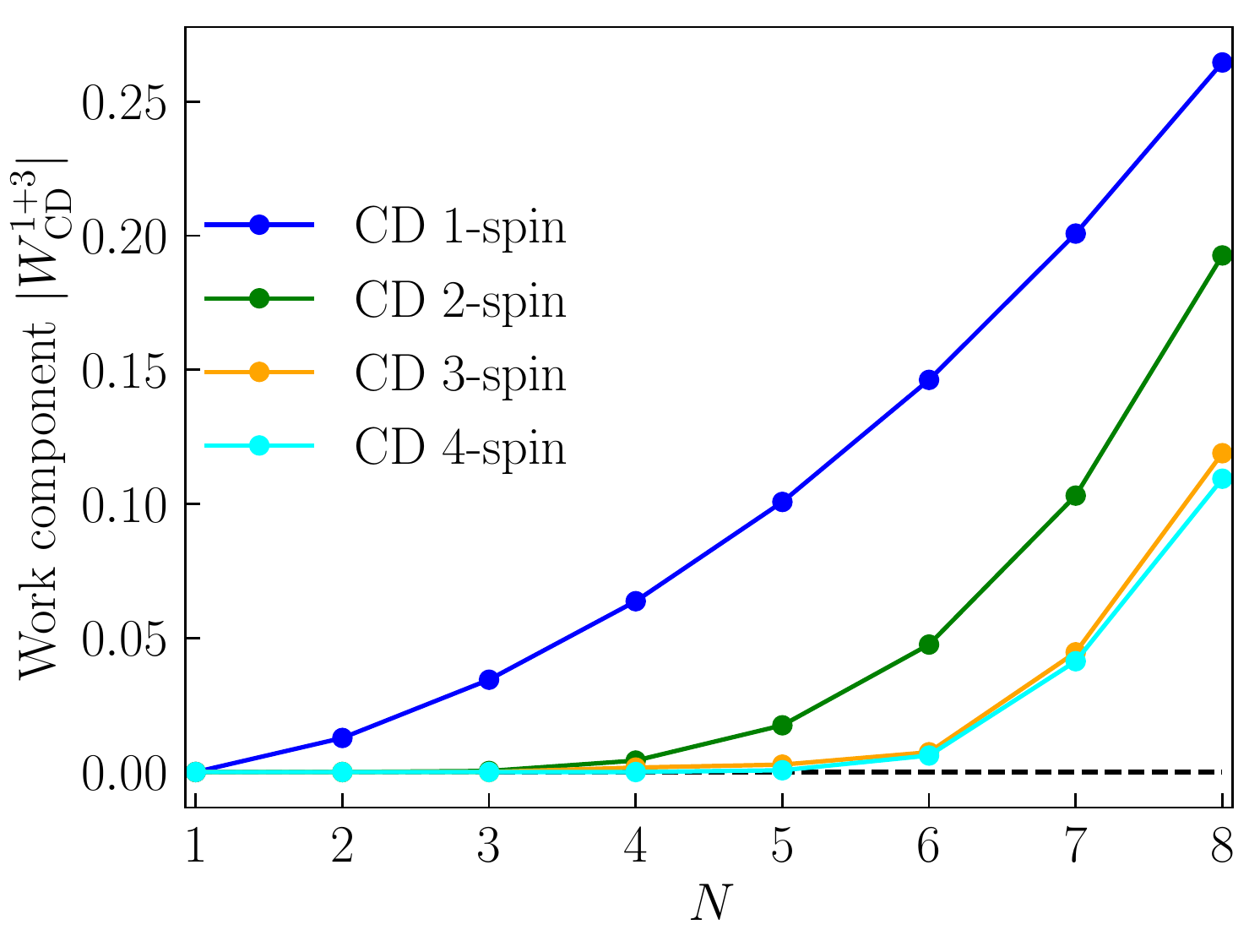}
  \caption{\textbf{Work component $\pmb{|W^{1+3}_\mathrm{CD}|}$ over system size $\pmb{N}$ for $\pmb{p \leq 4}$.} Absolute value of work component $W^{1+3}_\mathrm{CD}$ of the STA Hamiltonian [$H^*_\mathrm{STA}(t)$, Eq.~\eqref{eq_Hsta}] for different $p$-spin CD terms [$H^*_\mathrm{CD}(t)$, Eq.~\eqref{eq_Hcd}] over different system sizes $N$. Parameters: $\tau=\tau_1=\tau_3=1$. Other parameters as in Fig.~\ref{fig:fig2}.}
  \label{fig:fig3}
\end{figure}

Figure~\ref{fig:fig3} depicts the absolute value of the work component $W^{1+3}_\mathrm{CD}$ stemming from the external control device during one cycle for different system sizes $N$ up to $p=4$. We see that $W^{1+3}_\mathrm{CD}$ is zero as long as $p>N$. Namely, the CD Hamiltonian must comprise all kinds of interactions up to order $N$, i.e., involving all the spins in the chain. By contrast, for $N > p$, i.e., more spins $N$ in the WM than order of interaction $p$ in $H^*_\mathrm{CD}(t)$, we see that the absolute value of $W^{1+3}_\mathrm{CD}$ adopts a non-zero value which, according to Appendix~\ref{sec:C}, implies that the Hamiltonian $H^*_\mathrm{CD}(t)$ is \emph{not} exact anymore. We therefore conclude, that including a CD Hamiltonian with all combinations up to $N$-body terms leads to an exact expression when the WM contains $N$ spins which is consistent with Refs.~\cite{sels2017minimizing, claeys2019floquet}. 
These results encourage the goal to strive for an \emph{exact} CD drive. For the latter, the external control device fully assists the piston that optimally ``compresses'' and ``expands'' the quantum WM. 

\subsection{Dependence on cycle time}
\begin{figure}
  \centering
  \includegraphics[width=.95\columnwidth]{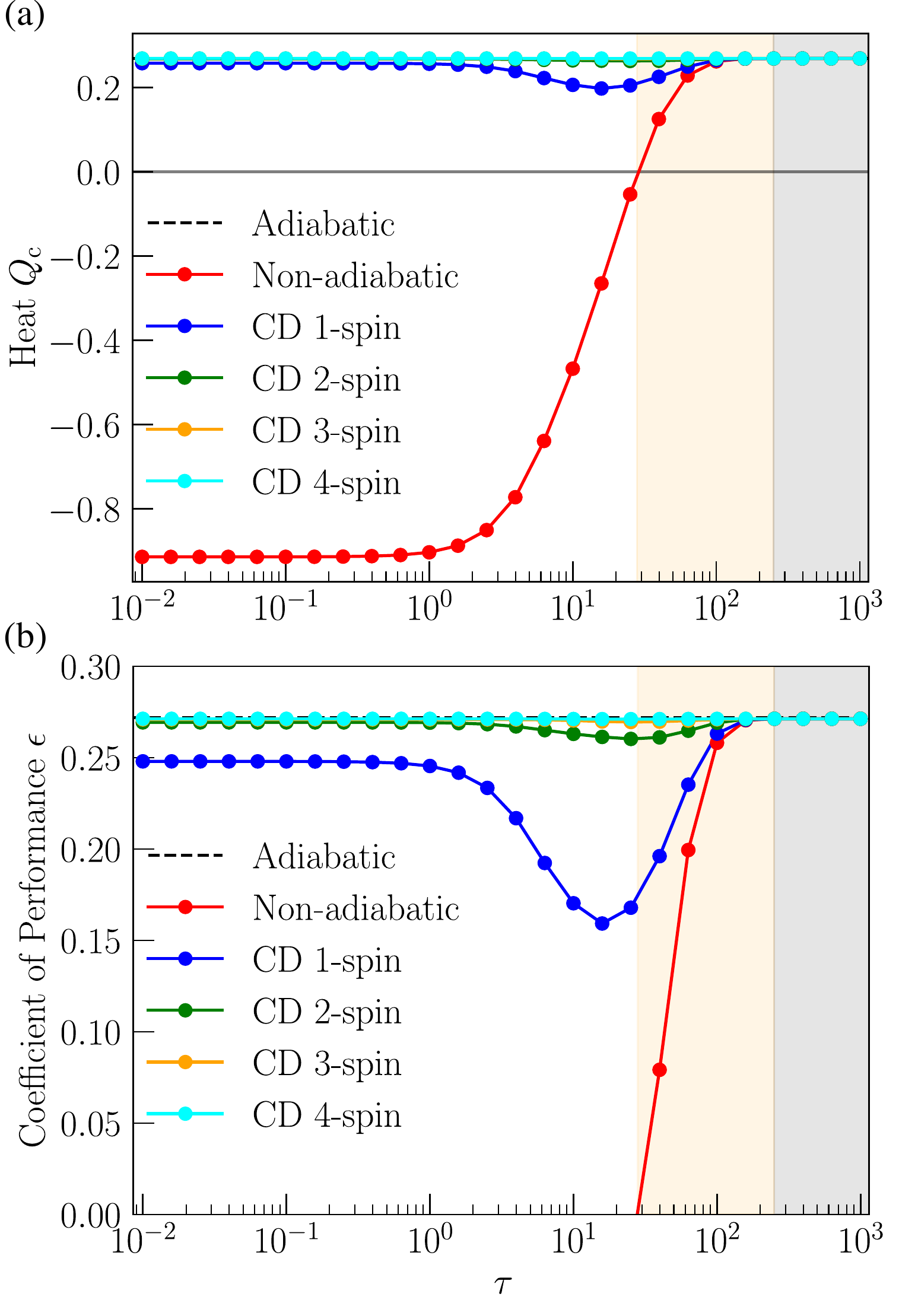}
  \caption{\textbf{Pumped heat and coefficient of performance.} (a)~Pumped heat $\Qc$ per cycle and (b)~CoP $\epsilon$ for (i)~non-adiabatic Hamiltonian [$H_0(t)$, Eq.~\eqref{eq_H0}] and (ii)~STA Hamiltonian [$H^*_\mathrm{STA}(t)$, Eq.~\eqref{eq_Hsta}] for different $p$-spin CD terms [$H^*_\mathrm{CD}(t)$, Eq.~\eqref{eq_Hcd}] over different isentropic stroke durations $\tau=\tau_1=\tau_3$ and system size $N=6$. 
Yellow-shaded area ($\tau \gtrsim 28$) depicts the refrigerator regime where $\Qc^{\mathrm{na}} > 0$ and the gray-shaded area ($\tau \geq 250$) the adiabatic regime where $\Qc^{\mathrm{na}} \approx \Qc^{\mathrm{ad}}$.
Other parameters as in Fig.~\ref{fig:fig2}.}
  \label{fig:fig4}
\end{figure}
We are further interested in the pumped heat $\Qc$ per cycle and the corresponding CoP for different cycle durations $\tau$.
Figure~\ref{fig:fig4}\textcolor{red}{(a)} shows the heat $\Qc$ extracted from the cold reservoir over one cycle for a system size $N=6$. 
For the quantum Otto cycle with non-adiabatic Hamiltonian $H_0(t)$, we see that heat $\Qc^{\mathrm{na}} > 0$ is pumped from the cold reservoir to the hot bath only for isentropic stroke durations $\tau = \tau_1 = \tau_3 \gtrsim 28$ (yellow-shaded area, refrigerator regime). For shorter isentropic stroke durations, the obtained states $\rho'_B$ and $\rho'_D$ at the end of each isentropic stroke (\cf Fig.~\ref{fig:fig1}) are so far away from the ideal adiabatic states $\rho_B$ and $\rho_D$ that the quantum Otto cycle ceases to describe a refrigerator. For large stroke durations, the refrigerator pumps the maximal possible cooling heat (gray-shaded area, adiabatic regime, $\tau \geq 250$ where $\Qc^{\mathrm{na}} \approx \Qc^{\mathrm{ad}}$).
\par
By contrast, the sped-up cycle with Hamiltonian $H^*_\mathrm{STA}(t)$ including $1$- up to $4$-spin CD terms $H^*_\mathrm{CD}(t)$ even pumps heat from the cold reservoir in the quench limit $\tau_1, \tau_3 \to 0$ where the cycle duration is dominated by thermalization. 
For increasing $p$, the cooling heat $\Qc$ increases to its maximally possible value $\Qc^{\mathrm{ad}}$ for all durations studied. Note, however, that $p=2$ appears to be efficient even for the intermediate regime where $p=1$ does not yield optimal results. This is in accordance with Fig.~\ref{fig:fig2}.
Note that for the adiabatic regime, the strength of the additional $H^*_\mathrm{CD}(t)$ converges to zero (as $\dot{\vartheta}(t) \varpropto 1/\tau$, \cf Appendix~\ref{sec:A}) and hence $\Qc \to \Qc^{\mathrm{ad}}$.

Figure~\ref{fig:fig4}\textcolor{red}{(b)} depicts the corresponding CoP $\epsilon$, Eq.~\eqref{eq_cop_sta}. For the sped-up cycle with $H^*_\mathrm{STA}(t)$, Eq.~\eqref{eq_Hsta}, we see that the higher $p$-spin terms we use for our CD Hamiltonian $H^*_\mathrm{CD}(t)$, Eq.~\eqref{eq_Hcd}, the larger the CoP becomes. 
Note that the CoP appears to be more sensitive to the value of $p$ than the pumped heat $\Qc$. Namely, converging to the optimal CoP $\epsilon^\mathrm{ad}$, Eq.~\eqref{eq_Cop_ad}, generally requires a larger $p$ than for the optimal exctracted heat $\Qc^{\mathrm{ad}}$ which is due to a non-zero $W^{1+3}_\mathrm{CD}$.

\section{Discussion and Outlook}\label{sec_dis}
In this work, we have presented a \emph{many-body} quantum Otto refrigerator that efficiently operates at \emph{finite-time}. The many-body quantum WM is modelled by an all-to-all connected Ising spin chain where the working parameter is parametrized via magnetic fields and interactions. In order to speed up the quantum Otto cycle, we apply an additional \emph{approximate} multi-spin CD Hamiltonian. The latter outperforms its non-adiabatic and traditional $1$-spin CD counterparts in pumped heat per cycle. The additional CD Hamiltonian contains $p$-spin terms with an odd number of $\sigma^y$ terms and $p \leq N$. For $p=N$, we obtain the exact CD Hamiltonian.
We provide an analytical proof that the work component stemming from the external control device is \emph{zero} if the CD Hamiltonian is \emph{exact}. In this case, the external control device fully assists the piston to pump heat from a cold to a hot reservoir and acts as a catalyst.
Note that while an exact CD protocol implies zero work contribution from the external control device over a cycle, the converse is not true. It is therefore not possible to find an exact CD Hamiltonian via minimizing this work component.

We numerically demonstrate an enhanced cooling power and coefficient of perfomance for the sped-up quantum Otto refrigerator with higher $p$-spin CD driving compared to its non-adiabatic and $1$-spin counterparts. Furthermore, we show that increasing $p$ improves the scaling behaviour in cooling power with system size. For the Otto cycle with single-body quantum WM, i.e., described via the Landau-Zener model, we find an analytical expression for the CoP that only depends on the applied magnetic field strengths. The analytical and numerical performance analysis reveals that quantum Otto refrigerators using WM with even all-to-all connectivity can be scaled up efficiently by applying multi-spin CD protocols while increasing the number of spins in the WM.

\par 

We note that several cost identifiers for the additional CD Hamiltonian have been introduced~\cite{abah2017energy, campbell2017tradeoff, zheng2016cost, abah2019shortcut, abah2018performance, cakmak2019spin, tobalina2019vanishing, abah2020shortcut}. We note, further, that these \emph{implementational} costs are conceptually different to the \emph{operational} costs described by the work $W^{1+3}_\mathrm{CD}$ stemming from the external control device~\cite{hartmann2020manybody}. In fact, in the case of exact CD driving we go on the optimal, i.e., adiabatic, path from points $A$ to $B$ and $C$ to $D$ in Fig.~\ref{fig:fig1}, respectively.
Although the operational cost will then be zero, i.e., $W^{1+3}_\mathrm{CD}=0$, there will still be a cost of implementing the additional CD Hamiltonian. In Appendix~\ref{sec:E} we have computed the strongly setup-dependent energetic cost of generating the additional magnetic fields of the CD Hamiltonian. However, although these energetic costs can be quantified, the actual implementation cost highly depends on the experimental setup and thus cannot be answered unambigously. 

\par

The results presented in this work disclose some important operational features.
This work reveals a trade-off between experimental feasibility of the additionally applied CD terms and vanishing work component stemming from the external control device. Namely, for the latter we need to apply non-local multi-spin CD terms (for example through additional laser fields) that are hard to implement in current experiments. On the other hand, applying only local $1$-spin CD terms may result in enhanced cooling power compared to its non-adiabatic counterpart, yet under the price of extensive usage of the external control device rather than the piston. This work further shows that we can efficiently speed up the Otto cycle even for increasing system sizes by applying sufficiently well approximated multi-spin CD Hamiltonians.
This work puts the results obtained in Ref.~\cite{hartmann2020manybody} into a wider context. In particular, this work significantly advances the ideas of the latter by providing a full assessment of experimental tuning parameters to consider and still being applicable for strongly site-dependent magnetic fields and interactions in a fully connected WM due to its increased complexity of external driving. While we have numerical evidence that $1$-spin CD driving is just able to speed up the cycle for weakly coupled spins in the WM, e.g. with nearest-neighbor interaction, the higher-spin CD driving even generates cooling heat for strongly coupled and site-dependent magnetic fields and interactions for various all-to-all connected WM.

\par 

For future research, we intend to study the robustness of the applied CD protocols with respect to external noise and possible cooperative effects~\cite{manatuly2019collectively}.
Furthermore, we aim at developing a many-body quantum refrigerator where work and heat exchanges occur simultaneously.

\section*{Acknowledgements}
We thank Adolfo del~Campo for valuable discussions.
W.\,N.\ acknowledges support from an ESQ fellowship of the Austrian Academy of Sciences (\"OAW).
V.\,M.\ acknowledges support from Science and Engineering Research Board (SERB) through Start-up Research Grant (Project No.SRG/2019/000411) and from Seed grant of IISER Berhampur. Work was supported by the Austrian Science Fund (FWF) through a START grant under Project No. Y1067-N27 and the SFB BeyondC Project No. F7108-N38, the Hauser-Raspe foundation, and the European Union's Horizon 2020 research and innovation program under grant agreement No. 817482. This material is based upon work supported by the Defense Advanced Research Projects Agency (DARPA) under Contract No. HR001120C0068. Any opinions, findings and conclusions or recommendations expressed in this material are those of the author(s) and do not necessarily reflect the views of DARPA.

\appendix
%\numberwithin{equation}{section}
\renewcommand{\theequation}{\thesection\arabic{equation}}
\setcounter{equation}{0}

\section{Protocols for magnetic fields and interactions}\label{sec:A}
For the many-body quantum WM in the text, the explicit time dependence of the magnetic field and interaction strengths for the non-adiabatic [$H_0(t)$, Eq.~\eqref{eq_H0}] and STA Hamiltonian [$H^*_\mathrm{STA}(t)$, Eq.~\eqref{eq_Hsta}] read
\begin{subequations}\label{app_fields_many-body}
  \begin{align}
    h_j(t)&=h_{j,\mathrm{i}}+(h_{j,\mathrm{f}}-h_{j,\mathrm{i}}) \vartheta(t),  \\
    b_j(t)&=b_{j,\mathrm{i}}+(b_{j,\mathrm{f}}-b_{j,\mathrm{i}}) \vartheta(t),  \\
    J_{jk}(t)&=J_{jk,\mathrm{i}}+(J_{jk,\mathrm{f}}-J_{jk, \mathrm{i}}) \vartheta(t)
    \label{eq_app_J}
  \end{align}
\end{subequations}
where 
\begin{equation}
\vartheta(t) \coloneqq \sin^2\left[\dfrac{\pi}{2}\sin^2\left(\dfrac{\pi t}{2 \tau_l}\right)\right]
\label{app_sweep_function}
\end{equation}
is the sweep function and $h_{j,\mathrm{i}}=h_j(t=0)$, $b_{j,\mathrm{i}}=b_j(t=0)$, $J_{jk,\mathrm{i}}=J_{jk}(t=0)$ and $h_{j,\mathrm{f}}=h_j(t=\tau_l)$, $b_{j,\mathrm{f}}=b_j(t=\tau_l)$, $J_{jk,\mathrm{f}}=J_{jk}(t=\tau_l)$ the initial and final values for each isentropic stroke $l \in \{1,3\}$ of duration $\tau_l$, respectively.
Its derivative with respect to time reads
\begin{equation}
\dot{\vartheta}(t)=\dfrac{\pi^2}{4\tau_l}\sin\left(\dfrac{\pi}{\tau_l}t\right)\sin\left[\pi \sin^2 \left(\dfrac{\pi}{2\tau_l}t \right)\right]
\label{app_derivative_vartheta}
\end{equation}
and is applied to the CD Hamiltonian, Eq.~\eqref{eq_Hcd}.

\setcounter{equation}{0}
\section{Multi-spin CD driving}\label{sec:B}
For the CD Hamiltonian $H_\mathrm{CD}(t)$, Eq.~\eqref{eq_Hcd} in the text, we apply an \emph{approximate} adiabatic gauge potential $\mathcal{A}^*_\vartheta$.
Here, we follow the variational method introduced in Ref.~\cite{sels2017minimizing}. 

We make an ansatz $\mathcal{A}^*_\vartheta$ with $p$-spin Pauli matrices and an odd number of $\sigma^y$ terms for the adiabatic gauge potential and calculate the Hermitian operator $G_\vartheta(\mathcal{A}^*_\vartheta) = \partial_\vartheta H_0 + i [\mathcal{A}^*_\vartheta, H_0]$. The goal is to minimize the operator distance $D^2(\mathcal{A}^*_\vartheta)= \mathrm{Tr}\{[G_\vartheta(\mathcal{A}_\vartheta) - G_\vartheta(\mathcal{A}^*_\vartheta)]^2\}$
between the exact, $\mathcal{A}_\vartheta$, and approximate, $\mathcal{A}^*_\vartheta$ adiabatic gauge potential.
This minimization is equivalent to minimizing the action 
\begin{equation}
S(\mathcal{A}^*_\vartheta) = \mathrm{Tr}[G^2_\vartheta(\mathcal{A}^*_\vartheta)]
\label{eq_appB_action}
\end{equation}
with respect to its parameters in front of every Pauli matrix term, i.e., $\delta \mathcal{S}(\mathcal{A}^*_\vartheta)/\delta \mathcal{A}^*_\vartheta=0$.

\par 

As an example, for $p=1$, i.e., 1-spin CD driving, we apply the ansatz $\mathcal{A}^*_\vartheta = \sum_{j=1}^N \alpha_j \sigma_j^y$ and calculate the Hermitian operator $G_\vartheta(\mathcal{A}^*_\vartheta)$ as well as the action $\mathcal{S}(\mathcal{A}^*_\vartheta)$. By minimizing the latter with respect to the coefficients $\alpha_j$, we find the optimal solution for each spin. 
For $p=2$, we apply the ansatz $\mathcal{A}^*_\vartheta = \sum_{j=1}^N \alpha_j \sigma_j^y + \sum_{k<j} \beta_{jk} (\sigma_j^y \sigma_k^z + \sigma_j^z \sigma_k^y) + \gamma_{jk} (\sigma_j^y \sigma_k^x + \sigma_j^x \sigma_k^y)$ and solve the corresponding action with respect to all coefficients $\alpha_j$, $\beta_{jk}$ and $\gamma_{jk}$. 
With this variational method, we can also include multi-spin terms, potentially up to $N$-body terms $\sigma_1^z \cdots \sigma_j^y \cdots \sigma_N^x$. We solve the optimal form of each coefficient numerically.

\setcounter{equation}{0}
\section{Proof of zero work component $W_\mathrm{CD}$ for exact CD driving}\label{sec:C}
Here, we provide a detailed proof to the statement that the work contribution~\cite{hartmann2020manybody} 
\begin{equation}
W_\mathrm{CD}^l = \int_0^{\tau_l} \mathrm{Tr} \left[\rho(t) \dot{H}_\mathrm{CD}(t)\right] dt
\end{equation}
stemming from the external control device for each isentropic stroke $l \in \{1,3 \}$ becomes \emph{zero} if the additionally applied CD Hamiltonian $H^*_\mathrm{CD}(t)$ is \emph{exact}.
The latter reads~\cite{berry2009transitionless}
\begin{equation}
H_\mathrm{CD}(t) = i \hbar \sum_n \vert\dot{n} \rangle \langle n \vert - \langle n \vert \dot{n} \rangle \vert n \rangle \langle n \vert
\end{equation}
where $\vert n \rangle$ denotes the instantaneous eigenstate at time $t$.
The density matrix $\rho(t)$ can be written as
\begin{equation}
\rho(t) = \sum_n a_n \vert n \rangle \langle n \vert
\end{equation}
where $a_n$ is a coefficient whose sum fulfils the normalization relation. Here, we have used that under exact CD driving the density matrix remains diagonal in the instantaneous eigenbasis~\cite{funo2017universal}.
\par
We now calculate the integral over the energy function
\begin{equation}
f(t) \coloneqq \mathrm{Tr} [ \rho(t) \dot{H}_\mathrm{CD}(t)]
\label{eq_function}
\end{equation}
for one isentropic stroke of duration $\tau$ [where $\dot{H}_\mathrm{CD}(t=0)=\dot{H}_\mathrm{CD}(t=\tau)=0$].
Employing the completeness condition $\sum_n \vert n \rangle \langle n \vert = \mathds{1}\nonumber$ of the eigenbasis, we receive the following relation
  \begin{align}
   &\sum_n \partial_t^2 (\vert n \rangle \langle n \vert) = \sum_n \vert \ddot{n} \rangle \langle n \vert + 2 \vert \dot{n} \rangle \langle \dot{n} \vert + \vert n \rangle \langle \ddot{n} \vert = 0 \nonumber \\
   &\Rightarrow \sum_n \vert \dot{n} \rangle \langle \dot{n} \vert =
 -\dfrac{1}{2} \sum_n \left(\vert n \rangle \langle \ddot{n} \vert + \vert \ddot{n} \rangle \langle n \vert \right).
\label{eq_trick1}
\end{align}
On the other hand, the normalization condition $ \langle n  \vert n \rangle = 1$ for each eigenstate separately leads to the following equations
\begin{subequations}\label{eq_tricks}
\begin{align}
	&\partial_t (\langle n \vert n  \rangle) = \langle \dot{n} \vert n \rangle + \langle n \vert \dot{n} \rangle = 0,
 \label{eq_trick2}\\
      &\langle \dot{n} \vert n \rangle = - \langle n \vert \dot{n} \rangle \nonumber \\ 
      &\Rightarrow 2 \langle n \vert \dot{n} \rangle = \langle n \vert \dot{n} \rangle + \langle n \vert \dot{n} \rangle = \langle n \vert \dot{n} \rangle - \langle \dot{n} \vert  n \rangle.
 \label{eq_trick3}
  \end{align}
\end{subequations}
Consequently, the time-derivative of the Hamiltonian $H_\mathrm{CD}(t)$ reads
\begin{align}
\dot{H}_\mathrm{CD}(t)&= i \hbar \sum_n \vert \ddot{n} \rangle \langle n \vert + \vert \dot{n} \rangle \langle \dot{n} \vert - \partial_t (\langle n \vert \dot{n} \rangle \vert n \rangle \langle n \vert) \nonumber \\
& \stackrel{\mathclap{\mathrm{(C5)}}}{=} \,i \hbar \sum_n \dfrac{1}{2} (\vert \ddot{n} \rangle \langle n \vert - \vert n \rangle \langle \ddot{n} \vert) - \partial_t (\langle n \vert \dot{n} \rangle) \vert n \rangle \langle n \vert \nonumber \\
&- \langle n \vert \dot{n} \rangle (\vert \dot{n} \rangle \langle n \vert + \vert n \rangle \langle \dot{n} \vert).
\end{align}
The function $f(t)=\mathrm{Tr}[\rho(t) \dot{H}_\mathrm{CD}(t)]$, Eq.~\eqref{eq_function}, can consequently be rewritten (using  $\mathrm{Tr}[\hat{H}]=\sum_n \langle n | \hat{H} | n \rangle$) as
\begin{widetext}
\begin{align}
f(t)&= i \hbar \sum_n a_n \left[\dfrac{1}{2} \left(\langle n \vert \ddot{n} \rangle - \langle \ddot{n} \vert n \rangle \right) - \partial_t (\langle n \vert \dot{n} \rangle) - \langle n \vert \dot{n} \rangle \left( \langle n \vert \dot{n} \rangle + \langle \dot{n} \vert n \rangle \right) \right] \nonumber \\
&\stackrel{\mathclap{\mathrm{(C6a)}}}{=} \;\dfrac{i \hbar}{2} \sum_n a_n \left[ \langle n \vert \ddot{n} \rangle - \langle \ddot{n} \vert n \rangle - 2 \partial_t (\langle n \vert \dot{n} \rangle) \right] \nonumber \\
&\stackrel{\mathclap{\mathrm{(C6b)}}}{=} \; \dfrac{i \hbar}{2} \sum_n a_n \left[ \langle n \vert \ddot{n} \rangle - \langle \ddot{n} \vert n \rangle - \partial_t (\langle n \vert \dot{n} \rangle - \langle \dot{n} \vert n \rangle) \right] \nonumber \\
&= \dfrac{i \hbar}{2} \sum_n a_n \left[ \langle n \vert \ddot{n} \rangle - \langle \ddot{n} \vert n \rangle - \langle \dot{n} \vert \dot{n} \rangle - \langle n \vert \ddot{n} \rangle + \langle \ddot{n} \vert n \rangle + \langle \dot{n} \vert \dot{n} \rangle \right] \nonumber \\
&= 0
\end{align}
\end{widetext}
which proves the statement above. 

\setcounter{equation}{0}
\section{Single-body working medium}\label{sec:D}
In analogy to Ref.~\cite{cakmak2019spin}, we derive the CoP of the quantum refrigerator with a single-body WM.

The CoP, Eq.~\eqref{eq_cop_sta}, in the text can be rewritten in terms of the energies $E_i$ with $i \in \{A,B,C,D \}$ at each point (\cf Fig.~\ref{fig:fig1}) of the cycle. It reads
\begin{equation}
\epsilon_\mathrm{LZ}= \dfrac{E_A - E_D}{E_B -E_A + E_D -E_C}= \dfrac{1}{\dfrac{E_B-E_C}{E_A-E_D}-1}
\label{AppD_eq_CoP}
\end{equation}
where the energies are 
\begin{align}
E_A &= - h_{\mathrm{x,i}} \tanh \left(\dfrac{h_{\mathrm{x,i}}}{\Tc}\right), \nonumber \\
E_B &= - b_{\mathrm{z,f}} \tanh \left(\dfrac{h_{\mathrm{x,i}}}{\Tc}\right), \nonumber \\
E_C &= - b_{\mathrm{z,f}} \tanh \left(\dfrac{b_{\mathrm{z,f}}}{\Th}\right), \nonumber \\
E_D &= - h_{\mathrm{x,i}} \tanh \left(\dfrac{b_{\mathrm{z,f}}}{\Th}\right),
\label{AppD_eq_energies} 
\end{align}
due to the uniform scaling of the energy levels in this two-level system after each stroke. Inserting the latter, Eq.~\eqref{AppD_eq_CoP} consequently reads
\begin{equation}
\epsilon_\mathrm{LZ} = \dfrac{1}{\dfrac{b_\mathrm{z,f}}{h_\mathrm{x,i}}-1}
\label{AppD_eq_CoP_rewritten}
\end{equation}
which is equivalent to Eq.~\eqref{eq_CoP_LZ} from the text.

\par

The pumped heat 
\begin{equation}
\Qc = E_A - E_D = h_\mathrm{x,i} \left[ \tanh \left( \dfrac{b_\mathrm{z,f}}{\Th} \right) - \tanh \left( \dfrac{h_\mathrm{x,i}}{\Tc} \right) \right] \geq 0
\end{equation}
is positive for the Otto cycle to describe a refrigerator.
From the latter and the fact that $\tanh$ is a monotonously increasing function, it follows that
\begin{equation}
\dfrac{b_\mathrm{z,f}}{h_\mathrm{x,i}} \geq \dfrac{\Th}{\Tc}.
\end{equation} 
Inserting this into Eq.~\eqref{AppD_eq_CoP_rewritten}, we finally obtain
\begin{equation}
\epsilon_\mathrm{LZ} = \dfrac{1}{\dfrac{b_\mathrm{z,f}}{h_\mathrm{x,i}}-1} \leq \dfrac{1}{\dfrac{\Th}{\Tc}-1}= \dfrac{\Tc}{\Th - \Tc}= \epsilon_\mathrm{C}
\end{equation}
with the Carnot CoP $\epsilon_\mathrm{C}$ as the upper bound.

\setcounter{equation}{0}
\section{Implementation cost for multi-spin CD driving}\label{sec:E}
In this section we will consider the cost of implementing the additional Hamiltonian $H_\mathrm{CD}(t)$, Eq.~\eqref{eq_Hcd} from the text. To this end, we will compute the energetic cost 
\begin{equation}
\langle H^j_\mathrm{CD} \rangle = \nu_{t,N} \int_0^\tau \mathrm{Tr}[H^\dagger_\mathrm{CD}(t) H_\mathrm{CD}(t)] dt
\label{App_E_eq_implementation_cost}
\end{equation}
of generating the additional magnetic fields of the CD Hamiltonian as defined in Ref.~\cite{zheng2016cost}, for each isentropic stroke $j$ and squared Frobenius norm of the CD Hamiltonian, i.e. $\Vert H^j_\mathrm{CD}(t) \Vert^2$. Here $\nu_{t,N}$ is an experimental setup-dependent expression that can possibly be highly complicated to be evaluated.

\begin{figure}[ht]
  \centering
  \includegraphics[width=.9\columnwidth]{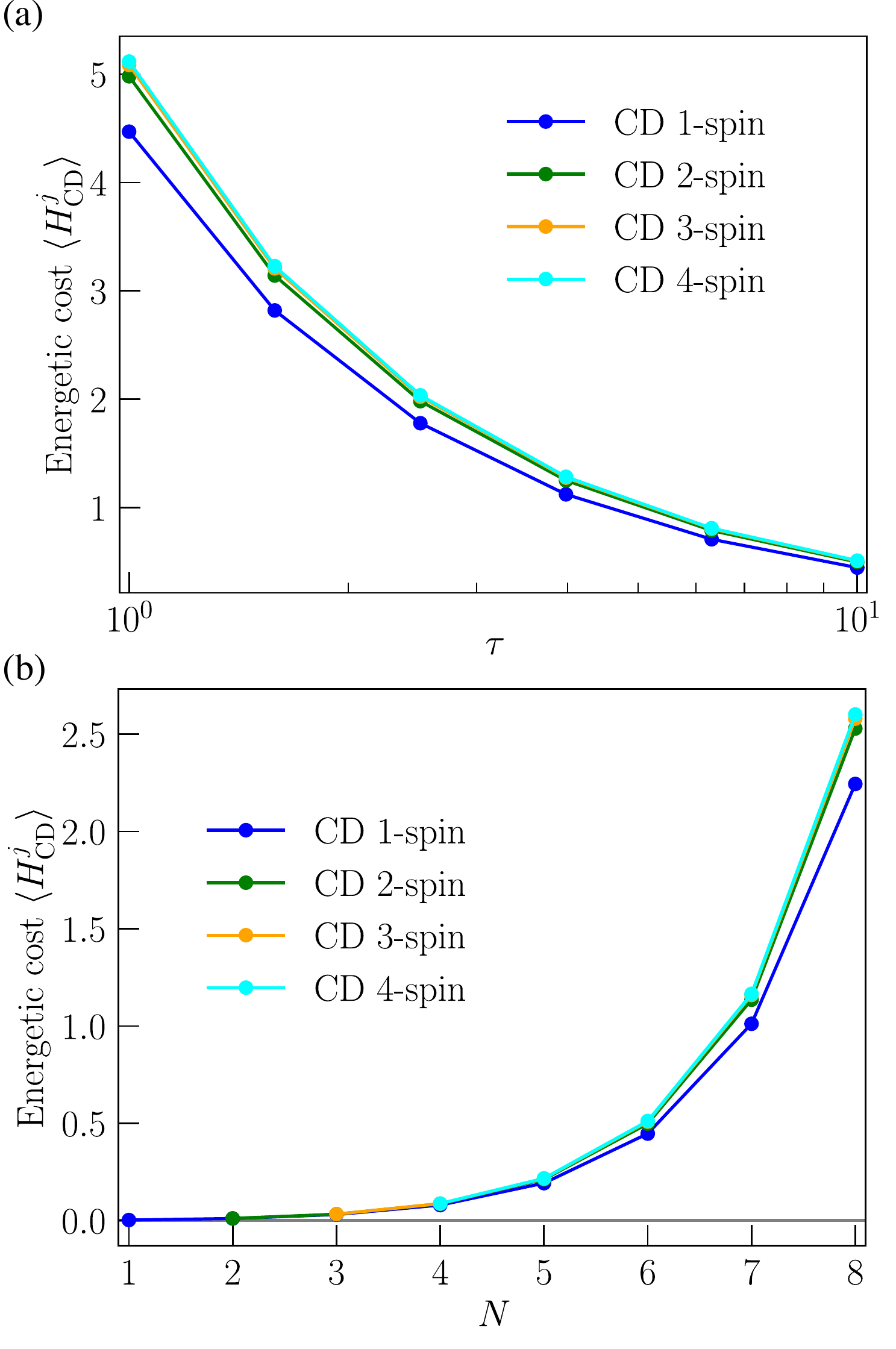}
  \caption{\textbf{Energetic costs.} Energetic cost $\langle H^j_\mathrm{CD} \rangle$, Eq.~\eqref{App_E_eq_implementation_cost}, for implementation of the CD Hamiltonian $H^j_\mathrm{CD}(t)$, Eq.~\eqref{eq_Hcd} from the text, during each isentropic stroke $j$ for (a)~different sweep durations $1 < \tau < 10$ and system size $N=6$ and (b)~different system sizes $N$ and sweep duration $\tau \approx 40$. Parameters: $\nu_{t,N}=0.01$ and other parameters as in Fig.~\ref{fig:fig2}.}
  \label{fig:fig5}
\end{figure}

Figure~\ref{fig:fig5} depicts this energetic cost for different $p$-spin CD driving for (a)~different sweep durations $1 < \tau < 10$ and system size $N=6$ and (b)~different system sizes $N$ and sweep duration $\tau \approx 40$ with same parameters as in Fig.~\ref{fig:fig2}.
For panel (a) we have studied the sweep times $1 < \tau < 10$ as the strength of the additional magnetic fields are in the same order as the original magnetic fields and interactions, respectively, which we consider to be highly interesting for experimental realization.

We see that the energetic costs decrease (increase) exponentially for increasing sweep duration $\tau$ (system size $N$). Although the difference in energetic cost between each $p$-spin CD Hamiltonian are rather small, in particular in between $2$-spin and $4$-spin CD driving, the actual implementation cost, i.e., implementing complicated non-local $N$-spin Hamiltonian rather than only local $1$-spin terms, may differ considerably.


\begin{thebibliography}{65}%
\makeatletter
\providecommand \@ifxundefined [1]{%
 \@ifx{#1\undefined}
}%
\providecommand \@ifnum [1]{%
 \ifnum #1\expandafter \@firstoftwo
 \else \expandafter \@secondoftwo
 \fi
}%
\providecommand \@ifx [1]{%
 \ifx #1\expandafter \@firstoftwo
 \else \expandafter \@secondoftwo
 \fi
}%
\providecommand \natexlab [1]{#1}%
\providecommand \enquote  [1]{#1}%
\providecommand \bibnamefont  [1]{#1}%
\providecommand \bibfnamefont [1]{#1}%
\providecommand \citenamefont [1]{#1}%
\providecommand \href@noop [0]{\@secondoftwo}%
\providecommand \href [0]{\begingroup \@sanitize@url \@href}%
\providecommand \@href[1]{\@@startlink{#1}\@@href}%
\providecommand \@@href[1]{\endgroup#1\@@endlink}%
\providecommand \@sanitize@url [0]{\catcode `\\12\catcode `\$12\catcode
  `\&12\catcode `\#12\catcode `\^12\catcode `\_12\catcode `\%12\relax}%
\providecommand \@@startlink[1]{}%
\providecommand \@@endlink[0]{}%
\providecommand \url  [0]{\begingroup\@sanitize@url \@url }%
\providecommand \@url [1]{\endgroup\@href {#1}{\urlprefix }}%
\providecommand \urlprefix  [0]{URL }%
\providecommand \Eprint [0]{\href }%
\providecommand \doibase [0]{http://dx.doi.org/}%
\providecommand \selectlanguage [0]{\@gobble}%
\providecommand \bibinfo  [0]{\@secondoftwo}%
\providecommand \bibfield  [0]{\@secondoftwo}%
\providecommand \translation [1]{[#1]}%
\providecommand \BibitemOpen [0]{}%
\providecommand \bibitemStop [0]{}%
\providecommand \bibitemNoStop [0]{.\EOS\space}%
\providecommand \EOS [0]{\spacefactor3000\relax}%
\providecommand \BibitemShut  [1]{\csname bibitem#1\endcsname}%
\let\auto@bib@innerbib\@empty
%</preamble>
\bibitem [{\citenamefont {\c{C}engel}\ and\ \citenamefont
  {Boles}(2015)}]{cengelbook}%
  \BibitemOpen
  \bibfield  {author} {\bibinfo {author} {\bibfnamefont {Y.~A.}\ \bibnamefont
  {\c{C}engel}}\ and\ \bibinfo {author} {\bibfnamefont {M.~A.}\ \bibnamefont
  {Boles}},\ }\href@noop {} {\emph {\bibinfo {title} {Thermodynamics: An
  Engineering Approach}}},\ \bibinfo {edition} {eighth}\ ed.\ (\bibinfo
  {publisher} {McGraw-Hill Education},\ \bibinfo {address} {New York},\
  \bibinfo {year} {2015})\BibitemShut {NoStop}%
\bibitem [{\citenamefont {Alicki}(1979)}]{alicki1979quantum}%
  \BibitemOpen
  \bibfield  {author} {\bibinfo {author} {\bibfnamefont {R.}~\bibnamefont
  {Alicki}},\ }\enquote {\bibinfo {title} {The quantum open system as a model
  of the heat engine},}\ \href {\doibase 10.1088/0305-4470/12/5/007} {\bibfield
   {journal} {\bibinfo  {journal} {J. Phys. A}\ }\textbf {\bibinfo {volume}
  {12}},\ \bibinfo {pages} {L103} (\bibinfo {year} {1979})}\BibitemShut
  {NoStop}%
\bibitem [{\citenamefont {Kosloff}(1984)}]{kosloff1984quantum}%
  \BibitemOpen
  \bibfield  {author} {\bibinfo {author} {\bibfnamefont {R.}~\bibnamefont
  {Kosloff}},\ }\enquote {\bibinfo {title} {A quantum mechanical open system as
  a model of a heat engine},}\ \href {\doibase 10.1063/1.446862} {\bibfield
  {journal} {\bibinfo  {journal} {J. Chem. Phys.}\ }\textbf {\bibinfo {volume}
  {80}},\ \bibinfo {pages} {1625} (\bibinfo {year} {1984})}\BibitemShut
  {NoStop}%
\bibitem [{\citenamefont {Kosloff}(2013)}]{kosloff2013quantum}%
  \BibitemOpen
  \bibfield  {author} {\bibinfo {author} {\bibfnamefont {R.}~\bibnamefont
  {Kosloff}},\ }\enquote {\bibinfo {title} {Quantum Thermodynamics: A Dynamical
  Viewpoint},}\ \href {\doibase 10.3390/e15062100} {\bibfield  {journal}
  {\bibinfo  {journal} {Entropy}\ }\textbf {\bibinfo {volume} {15}},\ \bibinfo
  {pages} {2100} (\bibinfo {year} {2013})}\BibitemShut {NoStop}%
\bibitem [{\citenamefont {Gelbwaser-Klimovsky}\ \emph
  {et~al.}(2015)\citenamefont {Gelbwaser-Klimovsky}, \citenamefont {Niedenzu},\
  and\ \citenamefont {Kurizki}}]{gelbwaser2015thermodynamics}%
  \BibitemOpen
  \bibfield  {author} {\bibinfo {author} {\bibfnamefont {D.}~\bibnamefont
  {Gelbwaser-Klimovsky}}, \bibinfo {author} {\bibfnamefont {W.}~\bibnamefont
  {Niedenzu}}, \ and\ \bibinfo {author} {\bibfnamefont {G.}~\bibnamefont
  {Kurizki}},\ }\enquote {\bibinfo {title} {Thermodynamics of Quantum Systems
  Under Dynamical Control},}\ \href {\doibase 10.1016/bs.aamop.2015.07.002}
  {\bibfield  {journal} {\bibinfo  {journal} {Adv. At. Mol. Opt. Phys.}\
  }\textbf {\bibinfo {volume} {64}},\ \bibinfo {pages} {329} (\bibinfo {year}
  {2015})}\BibitemShut {NoStop}%
\bibitem [{\citenamefont {Vinjanampathy}\ and\ \citenamefont
  {Anders}(2016)}]{vinjanampathy2016quantum}%
  \BibitemOpen
  \bibfield  {author} {\bibinfo {author} {\bibfnamefont {S.}~\bibnamefont
  {Vinjanampathy}}\ and\ \bibinfo {author} {\bibfnamefont {J.}~\bibnamefont
  {Anders}},\ }\enquote {\bibinfo {title} {Quantum thermodynamics},}\ \href
  {\doibase 10.1080/00107514.2016.1201896} {\bibfield  {journal} {\bibinfo
  {journal} {Contemp. Phys.}\ }\textbf {\bibinfo {volume} {57}},\ \bibinfo
  {pages} {1} (\bibinfo {year} {2016})}\BibitemShut {NoStop}%
\bibitem [{\citenamefont {Karimi}\ and\ \citenamefont
  {Pekola}(2016)}]{karimi2016otto}%
  \BibitemOpen
  \bibfield  {author} {\bibinfo {author} {\bibfnamefont {B.}~\bibnamefont
  {Karimi}}\ and\ \bibinfo {author} {\bibfnamefont {J.~P.}\ \bibnamefont
  {Pekola}},\ }\enquote {\bibinfo {title} {Otto refrigerator based on a
  superconducting qubit: Classical and quantum performance},}\ \href {\doibase
  10.1103/PhysRevB.94.184503} {\bibfield  {journal} {\bibinfo  {journal} {Phys.
  Rev. B}\ }\textbf {\bibinfo {volume} {94}},\ \bibinfo {pages} {184503}
  (\bibinfo {year} {2016})}\BibitemShut {NoStop}%
\bibitem [{\citenamefont {Kosloff}\ and\ \citenamefont
  {Rezek}(2017)}]{kosloff2017quantum}%
  \BibitemOpen
  \bibfield  {author} {\bibinfo {author} {\bibfnamefont {R.}~\bibnamefont
  {Kosloff}}\ and\ \bibinfo {author} {\bibfnamefont {Y.}~\bibnamefont
  {Rezek}},\ }\enquote {\bibinfo {title} {The Quantum Harmonic Otto Cycle},}\
  \href {\doibase 10.3390/e19040136} {\bibfield  {journal} {\bibinfo  {journal}
  {Entropy}\ }\textbf {\bibinfo {volume} {19}} (\bibinfo {year} {2017}),\
  10.3390/e19040136}\BibitemShut {NoStop}%
\bibitem [{\citenamefont {Binder}\ \emph {et~al.}(2019)\citenamefont {Binder},
  \citenamefont {Correa}, \citenamefont {Gogolin}, \citenamefont {Anders},\
  and\ \citenamefont {Adesso}}]{binder2019thermodynamicsbook}%
  \BibitemOpen
  \bibinfo {editor} {\bibfnamefont {F.}~\bibnamefont {Binder}}, \bibinfo
  {editor} {\bibfnamefont {L.~A.}\ \bibnamefont {Correa}}, \bibinfo {editor}
  {\bibfnamefont {C.}~\bibnamefont {Gogolin}}, \bibinfo {editor} {\bibfnamefont
  {J.}~\bibnamefont {Anders}}, \ and\ \bibinfo {editor} {\bibfnamefont
  {G.}~\bibnamefont {Adesso}},\ eds.,\ \href {\doibase
  10.1007/978-3-319-99046-0} {\emph {\bibinfo {title} {Thermodynamics in the
  Quantum Regime}}}\ (\bibinfo  {publisher} {Springer},\ \bibinfo {address}
  {Cham},\ \bibinfo {year} {2019})\BibitemShut {NoStop}%
\bibitem [{\citenamefont {Bhattacharjee}\ and\ \citenamefont
  {Dutta}(2020)}]{bhattacharjee2020quantum}%
  \BibitemOpen
  \bibfield  {author} {\bibinfo {author} {\bibfnamefont {S.}~\bibnamefont
  {Bhattacharjee}}\ and\ \bibinfo {author} {\bibfnamefont {A.}~\bibnamefont
  {Dutta}},\ }\href@noop {} {\enquote {\bibinfo {title} {Quantum thermal
  machines and batteries},}\ } (\bibinfo {year} {2020}),\ \Eprint
  {http://arxiv.org/abs/2008.07889} {arXiv:2008.07889 [quant-ph]} \BibitemShut
  {NoStop}%
\bibitem [{\citenamefont {Bernien}\ \emph {et~al.}(2017)\citenamefont
  {Bernien}, \citenamefont {Schwartz}, \citenamefont {Keesling}, \citenamefont
  {Levine}, \citenamefont {Omran}, \citenamefont {Pichler}, \citenamefont
  {Choi}, \citenamefont {Zibrov}, \citenamefont {Endres}, \citenamefont
  {Greiner}, \citenamefont {Vuleti\'{c}},\ and\ \citenamefont
  {Lukin}}]{bernien2017probing}%
  \BibitemOpen
  \bibfield  {author} {\bibinfo {author} {\bibfnamefont {H.}~\bibnamefont
  {Bernien}}, \bibinfo {author} {\bibfnamefont {S.}~\bibnamefont {Schwartz}},
  \bibinfo {author} {\bibfnamefont {A.}~\bibnamefont {Keesling}}, \bibinfo
  {author} {\bibfnamefont {H.}~\bibnamefont {Levine}}, \bibinfo {author}
  {\bibfnamefont {A.}~\bibnamefont {Omran}}, \bibinfo {author} {\bibfnamefont
  {H.}~\bibnamefont {Pichler}}, \bibinfo {author} {\bibfnamefont
  {S.}~\bibnamefont {Choi}}, \bibinfo {author} {\bibfnamefont {A.~S.}\
  \bibnamefont {Zibrov}}, \bibinfo {author} {\bibfnamefont {M.}~\bibnamefont
  {Endres}}, \bibinfo {author} {\bibfnamefont {M.}~\bibnamefont {Greiner}},
  \bibinfo {author} {\bibfnamefont {V.}~\bibnamefont {Vuleti\'{c}}}, \ and\
  \bibinfo {author} {\bibfnamefont {M.~D.}\ \bibnamefont {Lukin}},\ }\enquote
  {\bibinfo {title} {Probing many-body dynamics on a 51-atom quantum
  simulator},}\ \href {\doibase 10.1038/nature24622} {\bibfield  {journal}
  {\bibinfo  {journal} {Nature}\ }\textbf {\bibinfo {volume} {551}},\ \bibinfo
  {pages} {579} (\bibinfo {year} {2017})}\BibitemShut {NoStop}%
\bibitem [{\citenamefont {Choi}\ \emph {et~al.}(2016)\citenamefont {Choi},
  \citenamefont {Hild}, \citenamefont {Zeiher}, \citenamefont {Schau{\ss}},
  \citenamefont {Rubio-Abadal}, \citenamefont {Yefsah}, \citenamefont
  {Khemani}, \citenamefont {Huse}, \citenamefont {Bloch},\ and\ \citenamefont
  {Gross}}]{choi2016exploring}%
  \BibitemOpen
  \bibfield  {author} {\bibinfo {author} {\bibfnamefont {J.-y.}\ \bibnamefont
  {Choi}}, \bibinfo {author} {\bibfnamefont {S.}~\bibnamefont {Hild}}, \bibinfo
  {author} {\bibfnamefont {J.}~\bibnamefont {Zeiher}}, \bibinfo {author}
  {\bibfnamefont {P.}~\bibnamefont {Schau{\ss}}}, \bibinfo {author}
  {\bibfnamefont {A.}~\bibnamefont {Rubio-Abadal}}, \bibinfo {author}
  {\bibfnamefont {T.}~\bibnamefont {Yefsah}}, \bibinfo {author} {\bibfnamefont
  {V.}~\bibnamefont {Khemani}}, \bibinfo {author} {\bibfnamefont {D.~A.}\
  \bibnamefont {Huse}}, \bibinfo {author} {\bibfnamefont {I.}~\bibnamefont
  {Bloch}}, \ and\ \bibinfo {author} {\bibfnamefont {C.}~\bibnamefont
  {Gross}},\ }\enquote {\bibinfo {title} {Exploring the many-body localization
  transition in two dimensions},}\ \href {\doibase 10.1126/science.aaf8834}
  {\bibfield  {journal} {\bibinfo  {journal} {Science}\ }\textbf {\bibinfo
  {volume} {352}},\ \bibinfo {pages} {1547} (\bibinfo {year}
  {2016})}\BibitemShut {NoStop}%
\bibitem [{\citenamefont {Bordia}\ \emph {et~al.}(2017)\citenamefont {Bordia},
  \citenamefont {L\"uschen}, \citenamefont {Scherg}, \citenamefont
  {Gopalakrishnan}, \citenamefont {Knap}, \citenamefont {Schneider},\ and\
  \citenamefont {Bloch}}]{bordia2017probing}%
  \BibitemOpen
  \bibfield  {author} {\bibinfo {author} {\bibfnamefont {P.}~\bibnamefont
  {Bordia}}, \bibinfo {author} {\bibfnamefont {H.}~\bibnamefont {L\"uschen}},
  \bibinfo {author} {\bibfnamefont {S.}~\bibnamefont {Scherg}}, \bibinfo
  {author} {\bibfnamefont {S.}~\bibnamefont {Gopalakrishnan}}, \bibinfo
  {author} {\bibfnamefont {M.}~\bibnamefont {Knap}}, \bibinfo {author}
  {\bibfnamefont {U.}~\bibnamefont {Schneider}}, \ and\ \bibinfo {author}
  {\bibfnamefont {I.}~\bibnamefont {Bloch}},\ }\enquote {\bibinfo {title}
  {Probing Slow Relaxation and Many-Body Localization in Two-Dimensional
  Quasiperiodic Systems},}\ \href {\doibase 10.1103/PhysRevX.7.041047}
  {\bibfield  {journal} {\bibinfo  {journal} {Phys. Rev. X}\ }\textbf {\bibinfo
  {volume} {7}},\ \bibinfo {pages} {041047} (\bibinfo {year}
  {2017})}\BibitemShut {NoStop}%
\bibitem [{\citenamefont {Koski}\ \emph {et~al.}(2014)\citenamefont {Koski},
  \citenamefont {Maisi}, \citenamefont {Pekola},\ and\ \citenamefont
  {Averin}}]{koski2014experimental}%
  \BibitemOpen
  \bibfield  {author} {\bibinfo {author} {\bibfnamefont {J.~V.}\ \bibnamefont
  {Koski}}, \bibinfo {author} {\bibfnamefont {V.~F.}\ \bibnamefont {Maisi}},
  \bibinfo {author} {\bibfnamefont {J.~P.}\ \bibnamefont {Pekola}}, \ and\
  \bibinfo {author} {\bibfnamefont {D.~V.}\ \bibnamefont {Averin}},\ }\enquote
  {\bibinfo {title} {Experimental realization of a Szilard engine with a single
  electron},}\ \href {\doibase 10.1073/pnas.1406966111} {\bibfield  {journal}
  {\bibinfo  {journal} {Proc. Natl. Acad. Sci. USA}\ }\textbf {\bibinfo
  {volume} {111}},\ \bibinfo {pages} {13786} (\bibinfo {year}
  {2014})}\BibitemShut {NoStop}%
\bibitem [{\citenamefont {Ro{\ss}nagel}\ \emph {et~al.}(2016)\citenamefont
  {Ro{\ss}nagel}, \citenamefont {Dawkins}, \citenamefont {Tolazzi},
  \citenamefont {Abah}, \citenamefont {Lutz}, \citenamefont {Schmidt-Kaler},\
  and\ \citenamefont {Singer}}]{rossnagel2016single}%
  \BibitemOpen
  \bibfield  {author} {\bibinfo {author} {\bibfnamefont {J.}~\bibnamefont
  {Ro{\ss}nagel}}, \bibinfo {author} {\bibfnamefont {S.~T.}\ \bibnamefont
  {Dawkins}}, \bibinfo {author} {\bibfnamefont {K.~N.}\ \bibnamefont
  {Tolazzi}}, \bibinfo {author} {\bibfnamefont {O.}~\bibnamefont {Abah}},
  \bibinfo {author} {\bibfnamefont {E.}~\bibnamefont {Lutz}}, \bibinfo {author}
  {\bibfnamefont {F.}~\bibnamefont {Schmidt-Kaler}}, \ and\ \bibinfo {author}
  {\bibfnamefont {K.}~\bibnamefont {Singer}},\ }\enquote {\bibinfo {title} {A
  single-atom heat engine},}\ \href {\doibase 10.1126/science.aad6320}
  {\bibfield  {journal} {\bibinfo  {journal} {Science}\ }\textbf {\bibinfo
  {volume} {352}},\ \bibinfo {pages} {325} (\bibinfo {year}
  {2016})}\BibitemShut {NoStop}%
\bibitem [{\citenamefont {Klaers}\ \emph {et~al.}(2017)\citenamefont {Klaers},
  \citenamefont {Faelt}, \citenamefont {Imamoglu},\ and\ \citenamefont
  {Togan}}]{klaers2017squeezed}%
  \BibitemOpen
  \bibfield  {author} {\bibinfo {author} {\bibfnamefont {J.}~\bibnamefont
  {Klaers}}, \bibinfo {author} {\bibfnamefont {S.}~\bibnamefont {Faelt}},
  \bibinfo {author} {\bibfnamefont {A.}~\bibnamefont {Imamoglu}}, \ and\
  \bibinfo {author} {\bibfnamefont {E.}~\bibnamefont {Togan}},\ }\enquote
  {\bibinfo {title} {Squeezed Thermal Reservoirs as a Resource for a
  Nanomechanical Engine beyond the Carnot Limit},}\ \href {\doibase
  10.1103/PhysRevX.7.031044} {\bibfield  {journal} {\bibinfo  {journal} {Phys.
  Rev. X}\ }\textbf {\bibinfo {volume} {7}},\ \bibinfo {pages} {031044}
  (\bibinfo {year} {2017})}\BibitemShut {NoStop}%
\bibitem [{\citenamefont {Peterson}\ \emph {et~al.}(2019)\citenamefont
  {Peterson}, \citenamefont {Batalh\~ao}, \citenamefont {Herrera},
  \citenamefont {Souza}, \citenamefont {Sarthour}, \citenamefont {Oliveira},\
  and\ \citenamefont {Serra}}]{peterson2019experimental}%
  \BibitemOpen
  \bibfield  {author} {\bibinfo {author} {\bibfnamefont {J.~P.~S.}\
  \bibnamefont {Peterson}}, \bibinfo {author} {\bibfnamefont {T.~B.}\
  \bibnamefont {Batalh\~ao}}, \bibinfo {author} {\bibfnamefont
  {M.}~\bibnamefont {Herrera}}, \bibinfo {author} {\bibfnamefont {A.~M.}\
  \bibnamefont {Souza}}, \bibinfo {author} {\bibfnamefont {R.~S.}\ \bibnamefont
  {Sarthour}}, \bibinfo {author} {\bibfnamefont {I.~S.}\ \bibnamefont
  {Oliveira}}, \ and\ \bibinfo {author} {\bibfnamefont {R.~M.}\ \bibnamefont
  {Serra}},\ }\enquote {\bibinfo {title} {Experimental Characterization of a
  Spin Quantum Heat Engine},}\ \href {\doibase 10.1103/PhysRevLett.123.240601}
  {\bibfield  {journal} {\bibinfo  {journal} {Phys. Rev. Lett.}\ }\textbf
  {\bibinfo {volume} {123}},\ \bibinfo {pages} {240601} (\bibinfo {year}
  {2019})}\BibitemShut {NoStop}%
\bibitem [{\citenamefont {von Lindenfels}\ \emph {et~al.}(2019)\citenamefont
  {von Lindenfels}, \citenamefont {Gr\"ab}, \citenamefont {Schmiegelow},
  \citenamefont {Kaushal}, \citenamefont {Schulz}, \citenamefont {Mitchison},
  \citenamefont {Goold}, \citenamefont {Schmidt-Kaler},\ and\ \citenamefont
  {Poschinger}}]{vonlindenfels2019spin}%
  \BibitemOpen
  \bibfield  {author} {\bibinfo {author} {\bibfnamefont {D.}~\bibnamefont {von
  Lindenfels}}, \bibinfo {author} {\bibfnamefont {O.}~\bibnamefont {Gr\"ab}},
  \bibinfo {author} {\bibfnamefont {C.~T.}\ \bibnamefont {Schmiegelow}},
  \bibinfo {author} {\bibfnamefont {V.}~\bibnamefont {Kaushal}}, \bibinfo
  {author} {\bibfnamefont {J.}~\bibnamefont {Schulz}}, \bibinfo {author}
  {\bibfnamefont {M.~T.}\ \bibnamefont {Mitchison}}, \bibinfo {author}
  {\bibfnamefont {J.}~\bibnamefont {Goold}}, \bibinfo {author} {\bibfnamefont
  {F.}~\bibnamefont {Schmidt-Kaler}}, \ and\ \bibinfo {author} {\bibfnamefont
  {U.~G.}\ \bibnamefont {Poschinger}},\ }\enquote {\bibinfo {title} {Spin Heat
  Engine Coupled to a Harmonic-Oscillator Flywheel},}\ \href {\doibase
  10.1103/PhysRevLett.123.080602} {\bibfield  {journal} {\bibinfo  {journal}
  {Phys. Rev. Lett.}\ }\textbf {\bibinfo {volume} {123}},\ \bibinfo {pages}
  {080602} (\bibinfo {year} {2019})}\BibitemShut {NoStop}%
\bibitem [{\citenamefont {Klatzow}\ \emph {et~al.}(2019)\citenamefont
  {Klatzow}, \citenamefont {Becker}, \citenamefont {Ledingham}, \citenamefont
  {Weinzetl}, \citenamefont {Kaczmarek}, \citenamefont {Saunders},
  \citenamefont {Nunn}, \citenamefont {Walmsley}, \citenamefont {Uzdin},\ and\
  \citenamefont {Poem}}]{klatzow2019experimental}%
  \BibitemOpen
  \bibfield  {author} {\bibinfo {author} {\bibfnamefont {J.}~\bibnamefont
  {Klatzow}}, \bibinfo {author} {\bibfnamefont {J.~N.}\ \bibnamefont {Becker}},
  \bibinfo {author} {\bibfnamefont {P.~M.}\ \bibnamefont {Ledingham}}, \bibinfo
  {author} {\bibfnamefont {C.}~\bibnamefont {Weinzetl}}, \bibinfo {author}
  {\bibfnamefont {K.~T.}\ \bibnamefont {Kaczmarek}}, \bibinfo {author}
  {\bibfnamefont {D.~J.}\ \bibnamefont {Saunders}}, \bibinfo {author}
  {\bibfnamefont {J.}~\bibnamefont {Nunn}}, \bibinfo {author} {\bibfnamefont
  {I.~A.}\ \bibnamefont {Walmsley}}, \bibinfo {author} {\bibfnamefont
  {R.}~\bibnamefont {Uzdin}}, \ and\ \bibinfo {author} {\bibfnamefont
  {E.}~\bibnamefont {Poem}},\ }\enquote {\bibinfo {title} {Experimental
  Demonstration of Quantum Effects in the Operation of Microscopic Heat
  Engines},}\ \href {\doibase 10.1103/PhysRevLett.122.110601} {\bibfield
  {journal} {\bibinfo  {journal} {Phys. Rev. Lett.}\ }\textbf {\bibinfo
  {volume} {122}},\ \bibinfo {pages} {110601} (\bibinfo {year}
  {2019})}\BibitemShut {NoStop}%
\bibitem [{\citenamefont {Rezek}\ \emph {et~al.}(2009)\citenamefont {Rezek},
  \citenamefont {Salamon}, \citenamefont {Hoffmann},\ and\ \citenamefont
  {Kosloff}}]{rezek2009the}%
  \BibitemOpen
  \bibfield  {author} {\bibinfo {author} {\bibfnamefont {Y.}~\bibnamefont
  {Rezek}}, \bibinfo {author} {\bibfnamefont {P.}~\bibnamefont {Salamon}},
  \bibinfo {author} {\bibfnamefont {K.~H.}\ \bibnamefont {Hoffmann}}, \ and\
  \bibinfo {author} {\bibfnamefont {R.}~\bibnamefont {Kosloff}},\ }\enquote
  {\bibinfo {title} {The quantum refrigerator: The quest for absolute zero},}\
  \href {\doibase 10.1209/0295-5075/85/30008} {\bibfield  {journal} {\bibinfo
  {journal} {{EPL}}\ }\textbf {\bibinfo {volume} {85}},\
  \bibinfo {pages} {30008} (\bibinfo {year} {2009})}\BibitemShut {NoStop}%
\bibitem [{\citenamefont {Levy}\ and\ \citenamefont
  {Kosloff}(2012)}]{levy2012quantum}%
  \BibitemOpen
  \bibfield  {author} {\bibinfo {author} {\bibfnamefont {A.}~\bibnamefont
  {Levy}}\ and\ \bibinfo {author} {\bibfnamefont {R.}~\bibnamefont {Kosloff}},\
  }\enquote {\bibinfo {title} {Quantum Absorption Refrigerator},}\ \href
  {\doibase 10.1103/PhysRevLett.108.070604} {\bibfield  {journal} {\bibinfo
  {journal} {Phys. Rev. Lett.}\ }\textbf {\bibinfo {volume} {108}},\ \bibinfo
  {pages} {070604} (\bibinfo {year} {2012})}\BibitemShut {NoStop}%
\bibitem [{\citenamefont {Yuan}\ \emph {et~al.}(2014)\citenamefont {Yuan},
  \citenamefont {Wang}, \citenamefont {He}, \citenamefont {Ma},\ and\
  \citenamefont {Wang}}]{yuan2014coefficient}%
  \BibitemOpen
  \bibfield  {author} {\bibinfo {author} {\bibfnamefont {Y.}~\bibnamefont
  {Yuan}}, \bibinfo {author} {\bibfnamefont {R.}~\bibnamefont {Wang}}, \bibinfo
  {author} {\bibfnamefont {J.}~\bibnamefont {He}}, \bibinfo {author}
  {\bibfnamefont {Y.}~\bibnamefont {Ma}}, \ and\ \bibinfo {author}
  {\bibfnamefont {J.}~\bibnamefont {Wang}},\ }\enquote {\bibinfo {title}
  {Coefficient of performance under maximum $\ensuremath{\chi}$ criterion in a
  two-level atomic system as a refrigerator},}\ \href {\doibase
  10.1103/PhysRevE.90.052151} {\bibfield  {journal} {\bibinfo  {journal} {Phys.
  Rev. E}\ }\textbf {\bibinfo {volume} {90}},\ \bibinfo {pages} {052151}
  (\bibinfo {year} {2014})}\BibitemShut {NoStop}%
\bibitem [{\citenamefont {Long}\ and\ \citenamefont
  {Liu}(2015)}]{long2015performance}%
  \BibitemOpen
  \bibfield  {author} {\bibinfo {author} {\bibfnamefont {R.}~\bibnamefont
  {Long}}\ and\ \bibinfo {author} {\bibfnamefont {W.}~\bibnamefont {Liu}},\
  }\enquote {\bibinfo {title} {Performance of quantum Otto refrigerators with
  squeezing},}\ \href {\doibase 10.1103/PhysRevE.91.062137} {\bibfield
  {journal} {\bibinfo  {journal} {Phys. Rev. E}\ }\textbf {\bibinfo {volume}
  {91}},\ \bibinfo {pages} {062137} (\bibinfo {year} {2015})}\BibitemShut
  {NoStop}%
\bibitem [{\citenamefont {Abah}\ and\ \citenamefont
  {Lutz}(2016)}]{abah2016optimal}%
  \BibitemOpen
  \bibfield  {author} {\bibinfo {author} {\bibfnamefont {O.}~\bibnamefont
  {Abah}}\ and\ \bibinfo {author} {\bibfnamefont {E.}~\bibnamefont {Lutz}},\
  }\enquote {\bibinfo {title} {Optimal performance of a quantum Otto
  refrigerator},}\ \href {\doibase 10.1209/0295-5075/113/60002} {\bibfield
  {journal} {\bibinfo  {journal} {{EPL}}\ }\textbf
  {\bibinfo {volume} {113}},\ \bibinfo {pages} {60002} (\bibinfo {year}
  {2016})}\BibitemShut {NoStop}%
\bibitem [{\citenamefont {Niedenzu}\ \emph {et~al.}(2019)\citenamefont
  {Niedenzu}, \citenamefont {Mazets}, \citenamefont {Kurizki},\ and\
  \citenamefont {Jendrzejewski}}]{niedenzu2019quantized}%
  \BibitemOpen
  \bibfield  {author} {\bibinfo {author} {\bibfnamefont {W.}~\bibnamefont
  {Niedenzu}}, \bibinfo {author} {\bibfnamefont {I.}~\bibnamefont {Mazets}},
  \bibinfo {author} {\bibfnamefont {G.}~\bibnamefont {Kurizki}}, \ and\
  \bibinfo {author} {\bibfnamefont {F.}~\bibnamefont {Jendrzejewski}},\
  }\enquote {\bibinfo {title} {Quantized refrigerator for an atomic cloud},}\
  \href {\doibase 10.22331/q-2019-06-28-155} {\bibfield  {journal} {\bibinfo
  {journal} {{Quantum}}\ }\textbf {\bibinfo {volume} {3}},\ \bibinfo {pages}
  {155} (\bibinfo {year} {2019})}\BibitemShut {NoStop}%
\bibitem [{\citenamefont {Callen}(1985)}]{callen1985thermodynamics}%
  \BibitemOpen
  \bibfield  {author} {\bibinfo {author} {\bibfnamefont {H.~B.}\ \bibnamefont
  {Callen}},\ }\href@noop {} {\emph {\bibinfo {title} {Thermodynamics and an
  introduction to thermostatistics}}}\ (\bibinfo  {publisher} {Wiley},\
  \bibinfo {address} {New York},\ \bibinfo {year} {1985})\BibitemShut {NoStop}%
\bibitem [{\citenamefont {Arimondo}\ \emph {et~al.}(2013)\citenamefont
  {Arimondo}, \citenamefont {Berman},\ and\ \citenamefont
  {Lin}}]{arimondo2013chapter}%
  \BibitemOpen
  \bibinfo {editor} {\bibfnamefont {E.}~\bibnamefont {Arimondo}}, \bibinfo
  {editor} {\bibfnamefont {P.~R.}\ \bibnamefont {Berman}}, \ and\ \bibinfo
  {editor} {\bibfnamefont {C.~C.}\ \bibnamefont {Lin}},\ eds.,\ \href@noop {}
  {\emph {\bibinfo {title} {Advances in Atomic, Molecular, and Optical
  Physics}}},\ \bibinfo {series} {Adv. At. Mol. Opt. Phys.}, Vol.~\bibinfo
  {volume} {62}\ (\bibinfo  {publisher} {Academic Press},\ \bibinfo {year}
  {2013})\ pp.\ \bibinfo {pages} {117 -- 169}\BibitemShut {NoStop}%
\bibitem [{\citenamefont {del Campo}\ and\ \citenamefont
  {Sengupta}(2015)}]{delcampo2015controlling}%
  \BibitemOpen
  \bibfield  {author} {\bibinfo {author} {\bibfnamefont {A.}~\bibnamefont {del
  Campo}}\ and\ \bibinfo {author} {\bibfnamefont {K.}~\bibnamefont
  {Sengupta}},\ }\enquote {\bibinfo {title} {Controlling quantum critical
  dynamics of isolated systems},}\ \href {\doibase 10.1140/epjst/e2015-02350-4}
  {\bibfield  {journal} {\bibinfo  {journal} {Eur Phys J Spec Top}\ }\textbf {\bibinfo {volume} {224}},\ \bibinfo {pages} {189}
  (\bibinfo {year} {2015})}\BibitemShut {NoStop}%
\bibitem [{\citenamefont {del Campo}\ and\ \citenamefont
  {Kim}(2019)}]{delcampo2019focus}%
  \BibitemOpen
  \bibfield  {author} {\bibinfo {author} {\bibfnamefont {A.}~\bibnamefont {del
  Campo}}\ and\ \bibinfo {author} {\bibfnamefont {K.}~\bibnamefont {Kim}},\
  }\enquote {\bibinfo {title} {Focus on Shortcuts to Adiabaticity},}\ \href
  {\doibase 10.1088/1367-2630/ab1437} {\bibfield  {journal} {\bibinfo
  {journal} {New J. Phys.}\ }\textbf {\bibinfo {volume} {21}},\
  \bibinfo {pages} {050201} (\bibinfo {year} {2019})}\BibitemShut {NoStop}%
\bibitem [{\citenamefont {Gu\'ery-Odelin}\ \emph {et~al.}(2019)\citenamefont
  {Gu\'ery-Odelin}, \citenamefont {Ruschhaupt}, \citenamefont {Kiely},
  \citenamefont {Torrontegui}, \citenamefont {Mart\'{\i}nez-Garaot},\ and\
  \citenamefont {Muga}}]{guery-odelin2019shortcuts}%
  \BibitemOpen
  \bibfield  {author} {\bibinfo {author} {\bibfnamefont {D.}~\bibnamefont
  {Gu\'ery-Odelin}}, \bibinfo {author} {\bibfnamefont {A.}~\bibnamefont
  {Ruschhaupt}}, \bibinfo {author} {\bibfnamefont {A.}~\bibnamefont {Kiely}},
  \bibinfo {author} {\bibfnamefont {E.}~\bibnamefont {Torrontegui}}, \bibinfo
  {author} {\bibfnamefont {S.}~\bibnamefont {Mart\'{\i}nez-Garaot}}, \ and\
  \bibinfo {author} {\bibfnamefont {J.~G.}\ \bibnamefont {Muga}},\ }\enquote
  {\bibinfo {title} {Shortcuts to adiabaticity: Concepts, methods, and
  applications},}\ \href {\doibase 10.1103/RevModPhys.91.045001} {\bibfield
  {journal} {\bibinfo  {journal} {Rev. Mod. Phys.}\ }\textbf {\bibinfo {volume}
  {91}},\ \bibinfo {pages} {045001} (\bibinfo {year} {2019})}\BibitemShut
  {NoStop}%
\bibitem [{\citenamefont {Kosloff}\ and\ \citenamefont
  {Feldmann}(2002)}]{kosloff2002discrete}%
  \BibitemOpen
  \bibfield  {author} {\bibinfo {author} {\bibfnamefont {R.}~\bibnamefont
  {Kosloff}}\ and\ \bibinfo {author} {\bibfnamefont {T.}~\bibnamefont
  {Feldmann}},\ }\enquote {\bibinfo {title} {Discrete four-stroke quantum heat
  engine exploring the origin of friction},}\ \href {\doibase
  10.1103/PhysRevE.65.055102} {\bibfield  {journal} {\bibinfo  {journal} {Phys.
  Rev. E}\ }\textbf {\bibinfo {volume} {65}},\ \bibinfo {pages} {055102}
  (\bibinfo {year} {2002})}\BibitemShut {NoStop}%
\bibitem [{\citenamefont {Feldmann}\ and\ \citenamefont
  {Kosloff}(2003)}]{feldmann2003quantum}%
  \BibitemOpen
  \bibfield  {author} {\bibinfo {author} {\bibfnamefont {T.}~\bibnamefont
  {Feldmann}}\ and\ \bibinfo {author} {\bibfnamefont {R.}~\bibnamefont
  {Kosloff}},\ }\enquote {\bibinfo {title} {Quantum four-stroke heat engine:
  Thermodynamic observables in a model with intrinsic friction},}\ \href
  {\doibase 10.1103/PhysRevE.68.016101} {\bibfield  {journal} {\bibinfo
  {journal} {Phys. Rev. E}\ }\textbf {\bibinfo {volume} {68}},\ \bibinfo
  {pages} {016101} (\bibinfo {year} {2003})}\BibitemShut {NoStop}%
\bibitem [{\citenamefont {Feldmann}\ and\ \citenamefont
  {Kosloff}(2006)}]{feldmann2006quantum}%
  \BibitemOpen
  \bibfield  {author} {\bibinfo {author} {\bibfnamefont {T.}~\bibnamefont
  {Feldmann}}\ and\ \bibinfo {author} {\bibfnamefont {R.}~\bibnamefont
  {Kosloff}},\ }\enquote {\bibinfo {title} {Quantum lubrication: Suppression of
  friction in a first-principles four-stroke heat engine},}\ \href {\doibase
  10.1103/PhysRevE.73.025107} {\bibfield  {journal} {\bibinfo  {journal} {Phys.
  Rev. E}\ }\textbf {\bibinfo {volume} {73}},\ \bibinfo {pages} {025107}
  (\bibinfo {year} {2006})}\BibitemShut {NoStop}%
\bibitem [{\citenamefont {Demirplak}\ and\ \citenamefont
  {Rice}(2003)}]{demirplak2003adiabatic}%
  \BibitemOpen
  \bibfield  {author} {\bibinfo {author} {\bibfnamefont {M.}~\bibnamefont
  {Demirplak}}\ and\ \bibinfo {author} {\bibfnamefont {S.~A.}\ \bibnamefont
  {Rice}},\ }\enquote {\bibinfo {title} {Adiabatic Population Transfer with
  Control Fields},}\ \bibfield  {booktitle} {\emph {\bibinfo {booktitle} {J. Phys. Chem. A}},\ }\href {\doibase 10.1021/jp030708a}
  {\bibfield  {journal} {\bibinfo  {journal} {J. Phys. Chem. A}\ }\textbf {\bibinfo {volume} {107}},\ \bibinfo {pages} {9937} (\bibinfo
  {year} {2003})}\BibitemShut {NoStop}%
\bibitem [{\citenamefont
  {Takahashi}(2013{\natexlab{a}})}]{berry2009transitionless}%
  \BibitemOpen
  \bibfield  {author} {\bibinfo {author} {\bibfnamefont {K.}~\bibnamefont
  {Takahashi}},\ }\enquote {\bibinfo {title} {Transitionless quantum driving
  for spin systems},}\ \href {\doibase 10.1103/PhysRevE.87.062117} {\bibfield
  {journal} {\bibinfo  {journal} {Phys. Rev. E}\ }\textbf {\bibinfo {volume}
  {87}},\ \bibinfo {pages} {062117} (\bibinfo {year}
  {2013}{\natexlab{a}})}\BibitemShut {NoStop}%
\bibitem [{\citenamefont {Chen}\ \emph {et~al.}(2010)\citenamefont {Chen},
  \citenamefont {Ruschhaupt}, \citenamefont {Schmidt}, \citenamefont {del
  Campo}, \citenamefont {Gu\'ery-Odelin},\ and\ \citenamefont
  {Muga}}]{chen2010fast}%
  \BibitemOpen
  \bibfield  {author} {\bibinfo {author} {\bibfnamefont {X.}~\bibnamefont
  {Chen}}, \bibinfo {author} {\bibfnamefont {A.}~\bibnamefont {Ruschhaupt}},
  \bibinfo {author} {\bibfnamefont {S.}~\bibnamefont {Schmidt}}, \bibinfo
  {author} {\bibfnamefont {A.}~\bibnamefont {del Campo}}, \bibinfo {author}
  {\bibfnamefont {D.}~\bibnamefont {Gu\'ery-Odelin}}, \ and\ \bibinfo {author}
  {\bibfnamefont {J.~G.}\ \bibnamefont {Muga}},\ }\enquote {\bibinfo {title}
  {Fast Optimal Frictionless Atom Cooling in Harmonic Traps: Shortcut to
  Adiabaticity},}\ \href {\doibase 10.1103/PhysRevLett.104.063002} {\bibfield
  {journal} {\bibinfo  {journal} {Phys. Rev. Lett.}\ }\textbf {\bibinfo
  {volume} {104}},\ \bibinfo {pages} {063002} (\bibinfo {year}
  {2010})}\BibitemShut {NoStop}%
\bibitem [{\citenamefont {Chen}\ \emph {et~al.}(2011)\citenamefont {Chen},
  \citenamefont {Torrontegui},\ and\ \citenamefont {Muga}}]{chen2011lewis}%
  \BibitemOpen
  \bibfield  {author} {\bibinfo {author} {\bibfnamefont {X.}~\bibnamefont
  {Chen}}, \bibinfo {author} {\bibfnamefont {E.}~\bibnamefont {Torrontegui}}, \
  and\ \bibinfo {author} {\bibfnamefont {J.~G.}\ \bibnamefont {Muga}},\
  }\enquote {\bibinfo {title} {Lewis-Riesenfeld invariants and transitionless
  quantum driving},}\ \href {\doibase 10.1103/PhysRevA.83.062116} {\bibfield
  {journal} {\bibinfo  {journal} {Phys. Rev. A}\ }\textbf {\bibinfo {volume}
  {83}},\ \bibinfo {pages} {062116} (\bibinfo {year} {2011})}\BibitemShut
  {NoStop}%
\bibitem [{\citenamefont
  {Takahashi}(2013{\natexlab{b}})}]{takahashi2013transitionless}%
  \BibitemOpen
  \bibfield  {author} {\bibinfo {author} {\bibfnamefont {K.}~\bibnamefont
  {Takahashi}},\ }\enquote {\bibinfo {title} {Transitionless quantum driving
  for spin systems},}\ \href {\doibase 10.1103/PhysRevE.87.062117} {\bibfield
  {journal} {\bibinfo  {journal} {Phys. Rev. E}\ }\textbf {\bibinfo {volume}
  {87}},\ \bibinfo {pages} {062117} (\bibinfo {year}
  {2013}{\natexlab{b}})}\BibitemShut {NoStop}%
\bibitem [{\citenamefont {Jarzynski}(2013)}]{jarzynski2013generating}%
  \BibitemOpen
  \bibfield  {author} {\bibinfo {author} {\bibfnamefont {C.}~\bibnamefont
  {Jarzynski}},\ }\enquote {\bibinfo {title} {Generating shortcuts to
  adiabaticity in quantum and classical dynamics},}\ \href {\doibase
  10.1103/PhysRevA.88.040101} {\bibfield  {journal} {\bibinfo  {journal} {Phys.
  Rev. A}\ }\textbf {\bibinfo {volume} {88}},\ \bibinfo {pages} {040101}
  (\bibinfo {year} {2013})}\BibitemShut {NoStop}%
\bibitem [{\citenamefont {del Campo}(2013)}]{delcampo2013shortcuts}%
  \BibitemOpen
  \bibfield  {author} {\bibinfo {author} {\bibfnamefont {A.}~\bibnamefont {del
  Campo}},\ }\enquote {\bibinfo {title} {Shortcuts to Adiabaticity by
  Counterdiabatic Driving},}\ \href {\doibase 10.1103/PhysRevLett.111.100502}
  {\bibfield  {journal} {\bibinfo  {journal} {Phys. Rev. Lett.}\ }\textbf
  {\bibinfo {volume} {111}},\ \bibinfo {pages} {100502} (\bibinfo {year}
  {2013})}\BibitemShut {NoStop}%
\bibitem [{\citenamefont {Damski}(2014)}]{damski2014counterdiabatic}%
  \BibitemOpen
  \bibfield  {author} {\bibinfo {author} {\bibfnamefont {B.}~\bibnamefont
  {Damski}},\ }\enquote {\bibinfo {title} {Counterdiabatic driving of the
  quantum Ising model},}\ \href {\doibase 10.1088/1742-5468/2014/12/p12019}
  {\bibfield  {journal} {\bibinfo  {journal} {J. Stat. Mech. Theory Exp.}\ }\textbf {\bibinfo {volume} {2014}},\ \bibinfo
  {pages} {P12019} (\bibinfo {year} {2014})}\BibitemShut {NoStop}%
\bibitem [{\citenamefont {Sels}\ and\ \citenamefont
  {Polkovnikov}(2017)}]{sels2017minimizing}%
  \BibitemOpen
  \bibfield  {author} {\bibinfo {author} {\bibfnamefont {D.}~\bibnamefont
  {Sels}}\ and\ \bibinfo {author} {\bibfnamefont {A.}~\bibnamefont
  {Polkovnikov}},\ }\enquote {\bibinfo {title} {Minimizing irreversible losses
  in quantum systems by local counterdiabatic driving},}\ \href {\doibase
  10.1073/pnas.1619826114} {\bibfield  {journal} {\bibinfo  {journal} {Proc.
  Natl. Acad. Sci. USA}\ }\textbf {\bibinfo {volume} {114}},\ \bibinfo {pages}
  {E3909} (\bibinfo {year} {2017})}\BibitemShut {NoStop}%
\bibitem [{\citenamefont {Claeys}\ \emph {et~al.}(2019)\citenamefont {Claeys},
  \citenamefont {Pandey}, \citenamefont {Sels},\ and\ \citenamefont
  {Polkovnikov}}]{claeys2019floquet}%
  \BibitemOpen
  \bibfield  {author} {\bibinfo {author} {\bibfnamefont {P.~W.}\ \bibnamefont
  {Claeys}}, \bibinfo {author} {\bibfnamefont {M.}~\bibnamefont {Pandey}},
  \bibinfo {author} {\bibfnamefont {D.}~\bibnamefont {Sels}}, \ and\ \bibinfo
  {author} {\bibfnamefont {A.}~\bibnamefont {Polkovnikov}},\ }\enquote
  {\bibinfo {title} {Floquet-Engineering Counterdiabatic Protocols in Quantum
  Many-Body Systems},}\ \href {\doibase 10.1103/PhysRevLett.123.090602}
  {\bibfield  {journal} {\bibinfo  {journal} {Phys. Rev. Lett.}\ }\textbf
  {\bibinfo {volume} {123}},\ \bibinfo {pages} {090602} (\bibinfo {year}
  {2019})}\BibitemShut {NoStop}%
\bibitem [{\citenamefont {Hartmann}\ and\ \citenamefont
  {Lechner}(2019)}]{hartmann2019rapid}%
  \BibitemOpen
  \bibfield  {author} {\bibinfo {author} {\bibfnamefont {A.}~\bibnamefont
  {Hartmann}}\ and\ \bibinfo {author} {\bibfnamefont {W.}~\bibnamefont
  {Lechner}},\ }\enquote {\bibinfo {title} {Rapid counter-diabatic sweeps in
  lattice gauge adiabatic quantum computing},}\ \href {\doibase
  10.1088/1367-2630/ab14a0} {\bibfield  {journal} {\bibinfo  {journal} {New J.
  Phys.}\ }\textbf {\bibinfo {volume} {21}},\ \bibinfo {pages} {043025}
  (\bibinfo {year} {2019})}\BibitemShut {NoStop}%
\bibitem [{\citenamefont {Campo}\ \emph {et~al.}(2014)\citenamefont {Campo},
  \citenamefont {Goold},\ and\ \citenamefont {Paternostro}}]{delcampo2014more}%
  \BibitemOpen
  \bibfield  {author} {\bibinfo {author} {\bibfnamefont {A.~d.}\ \bibnamefont
  {Campo}}, \bibinfo {author} {\bibfnamefont {J.}~\bibnamefont {Goold}}, \ and\
  \bibinfo {author} {\bibfnamefont {M.}~\bibnamefont {Paternostro}},\ }\enquote
  {\bibinfo {title} {More bang for your buck: Super-adiabatic quantum
  engines},}\ \href {https://doi.org/10.1038/srep06208} {\bibfield  {journal}
  {\bibinfo  {journal} {Sci. Rep.}\ }\textbf {\bibinfo {volume} {4}},\
  \bibinfo {pages} {6208 EP } (\bibinfo {year} {2014})}\BibitemShut {NoStop}%
\bibitem [{\citenamefont {Abah}\ and\ \citenamefont
  {Lutz}(2018)}]{abah2018performance}%
  \BibitemOpen
  \bibfield  {author} {\bibinfo {author} {\bibfnamefont {O.}~\bibnamefont
  {Abah}}\ and\ \bibinfo {author} {\bibfnamefont {E.}~\bibnamefont {Lutz}},\
  }\enquote {\bibinfo {title} {Performance of shortcut-to-adiabaticity quantum
  engines},}\ \href {\doibase 10.1103/PhysRevE.98.032121} {\bibfield  {journal}
  {\bibinfo  {journal} {Phys. Rev. E}\ }\textbf {\bibinfo {volume} {98}},\
  \bibinfo {pages} {032121} (\bibinfo {year} {2018})}\BibitemShut {NoStop}%
\bibitem [{\citenamefont {Abah}\ and\ \citenamefont
  {Paternostro}(2019)}]{abah2019shortcut}%
  \BibitemOpen
  \bibfield  {author} {\bibinfo {author} {\bibfnamefont {O.}~\bibnamefont
  {Abah}}\ and\ \bibinfo {author} {\bibfnamefont {M.}~\bibnamefont
  {Paternostro}},\ }\enquote {\bibinfo {title} {Shortcut-to-adiabaticity Otto
  engine: A twist to finite-time thermodynamics},}\ \href {\doibase
  10.1103/PhysRevE.99.022110} {\bibfield  {journal} {\bibinfo  {journal} {Phys.
  Rev. E}\ }\textbf {\bibinfo {volume} {99}},\ \bibinfo {pages} {022110}
  (\bibinfo {year} {2019})}\BibitemShut {NoStop}%
\bibitem [{\citenamefont {Dupays}\ \emph {et~al.}(2020)\citenamefont {Dupays},
  \citenamefont {Egusquiza}, \citenamefont {del Campo},\ and\ \citenamefont
  {Chenu}}]{dupays2020superadiabatic}%
  \BibitemOpen
  \bibfield  {author} {\bibinfo {author} {\bibfnamefont {L.}~\bibnamefont
  {Dupays}}, \bibinfo {author} {\bibfnamefont {I.~L.}\ \bibnamefont
  {Egusquiza}}, \bibinfo {author} {\bibfnamefont {A.}~\bibnamefont {del
  Campo}}, \ and\ \bibinfo {author} {\bibfnamefont {A.}~\bibnamefont {Chenu}},\
  }\enquote {\bibinfo {title} {Superadiabatic thermalization of a quantum
  oscillator by engineered dephasing},}\ \href {\doibase
  10.1103/PhysRevResearch.2.033178} {\bibfield  {journal} {\bibinfo  {journal}
  {Phys. Rev. Research}\ }\textbf {\bibinfo {volume} {2}},\ \bibinfo {pages}
  {033178} (\bibinfo {year} {2020})}\BibitemShut {NoStop}%
\bibitem [{\citenamefont {\c{C}akmak}\ and\ \citenamefont {M\"ustecapl\ifmmode
  \imath \else \i \fi{}o\ifmmode~\breve{g}\else
  \u{g}\fi{}lu}(2019)}]{cakmak2019spin}%
  \BibitemOpen
  \bibfield  {author} {\bibinfo {author} {\bibfnamefont {B.}~\bibnamefont
  {\c{C}akmak}}\ and\ \bibinfo {author} {\bibfnamefont {{\"O}.~E.}\
  \bibnamefont {M\"ustecapl\ifmmode \imath \else \i
  \fi{}o\ifmmode~\breve{g}\else \u{g}\fi{}lu}},\ }\enquote {\bibinfo {title}
  {Spin quantum heat engines with shortcuts to adiabaticity},}\ \href {\doibase
  10.1103/PhysRevE.99.032108} {\bibfield  {journal} {\bibinfo  {journal} {Phys.
  Rev. E}\ }\textbf {\bibinfo {volume} {99}},\ \bibinfo {pages} {032108}
  (\bibinfo {year} {2019})}\BibitemShut {NoStop}%
\bibitem [{\citenamefont {Funo}\ \emph {et~al.}(2019)\citenamefont {Funo},
  \citenamefont {Lambert}, \citenamefont {Karimi}, \citenamefont {Pekola},
  \citenamefont {Masuyama},\ and\ \citenamefont {Nori}}]{funo2019speeding}%
  \BibitemOpen
  \bibfield  {author} {\bibinfo {author} {\bibfnamefont {K.}~\bibnamefont
  {Funo}}, \bibinfo {author} {\bibfnamefont {N.}~\bibnamefont {Lambert}},
  \bibinfo {author} {\bibfnamefont {B.}~\bibnamefont {Karimi}}, \bibinfo
  {author} {\bibfnamefont {J.~P.}\ \bibnamefont {Pekola}}, \bibinfo {author}
  {\bibfnamefont {Y.}~\bibnamefont {Masuyama}}, \ and\ \bibinfo {author}
  {\bibfnamefont {F.}~\bibnamefont {Nori}},\ }\enquote {\bibinfo {title}
  {Speeding up a quantum refrigerator via counterdiabatic driving},}\ \href
  {\doibase 10.1103/PhysRevB.100.035407} {\bibfield  {journal} {\bibinfo
  {journal} {Phys. Rev. B}\ }\textbf {\bibinfo {volume} {100}},\ \bibinfo
  {pages} {035407} (\bibinfo {year} {2019})}\BibitemShut {NoStop}%
\bibitem [{\citenamefont {Abah}\ \emph {et~al.}(2020)\citenamefont {Abah},
  \citenamefont {Paternostro},\ and\ \citenamefont {Lutz}}]{abah2020shortcut}%
  \BibitemOpen
  \bibfield  {author} {\bibinfo {author} {\bibfnamefont {O.}~\bibnamefont
  {Abah}}, \bibinfo {author} {\bibfnamefont {M.}~\bibnamefont {Paternostro}}, \
  and\ \bibinfo {author} {\bibfnamefont {E.}~\bibnamefont {Lutz}},\ }\enquote
  {\bibinfo {title} {Shortcut-to-adiabaticity quantum Otto refrigerator},}\
  \href {\doibase 10.1103/PhysRevResearch.2.023120} {\bibfield  {journal}
  {\bibinfo  {journal} {Phys. Rev. Research}\ }\textbf {\bibinfo {volume}
  {2}},\ \bibinfo {pages} {023120} (\bibinfo {year} {2020})}\BibitemShut
  {NoStop}%
\bibitem [{\citenamefont {Hartmann}\ \emph {et~al.}(2020)\citenamefont
  {Hartmann}, \citenamefont {Mukherjee}, \citenamefont {Niedenzu},\ and\
  \citenamefont {Lechner}}]{hartmann2020manybody}%
  \BibitemOpen
  \bibfield  {author} {\bibinfo {author} {\bibfnamefont {A.}~\bibnamefont
  {Hartmann}}, \bibinfo {author} {\bibfnamefont {V.}~\bibnamefont {Mukherjee}},
  \bibinfo {author} {\bibfnamefont {W.}~\bibnamefont {Niedenzu}}, \ and\
  \bibinfo {author} {\bibfnamefont {W.}~\bibnamefont {Lechner}},\ }\enquote
  {\bibinfo {title} {Many-body quantum heat engines with shortcuts to
  adiabaticity},}\ \href {\doibase 10.1103/PhysRevResearch.2.023145} {\bibfield
   {journal} {\bibinfo  {journal} {Phys. Rev. Research}\ }\textbf {\bibinfo
  {volume} {2}},\ \bibinfo {pages} {023145} (\bibinfo {year}
  {2020})}\BibitemShut {NoStop}%
  \bibitem [{\citenamefont {Cirac}\ and\ \citenamefont
  {Zoller}(2012)}]{cirac2012goals}%
  \BibitemOpen
  \bibfield  {author} {\bibinfo {author} {\bibfnamefont {J.~I.}\ \bibnamefont
  {Cirac}}\ and\ \bibinfo {author} {\bibfnamefont {P.}~\bibnamefont {Zoller}},\
  }\enquote {\bibinfo {title} {Goals and opportunities in quantum
  simulation},}\ \href {http://dx.doi.org/10.1038/nphys2275} {\bibfield
  {journal} {\bibinfo  {journal} {Nat. Phys.}\ }\textbf {\bibinfo {volume}
  {8}},\ \bibinfo {pages} {264} (\bibinfo {year} {2012})}\BibitemShut {NoStop}%
\bibitem [{\citenamefont {Georgescu}\ \emph {et~al.}(2014)\citenamefont
  {Georgescu}, \citenamefont {Ashhab},\ and\ \citenamefont
  {Nori}}]{georgescu2014quantum}%
  \BibitemOpen
  \bibfield  {author} {\bibinfo {author} {\bibfnamefont {I.~M.}\ \bibnamefont
  {Georgescu}}, \bibinfo {author} {\bibfnamefont {S.}~\bibnamefont {Ashhab}}, \
  and\ \bibinfo {author} {\bibfnamefont {F.}~\bibnamefont {Nori}},\ }\enquote
  {\bibinfo {title} {Quantum simulation},}\ \href {\doibase
  10.1103/RevModPhys.86.153} {\bibfield  {journal} {\bibinfo  {journal} {Rev.
  Mod. Phys.}\ }\textbf {\bibinfo {volume} {86}},\ \bibinfo {pages} {153}
  (\bibinfo {year} {2014})}\BibitemShut {NoStop}%
\bibitem [{\citenamefont {Lucas}(2014)}]{LucasAnnealing}%
  \BibitemOpen
  \bibfield  {author} {\bibinfo {author} {\bibfnamefont {A.}~\bibnamefont
  {Lucas}},\ }\enquote {\bibinfo {title} {Ising formulations of many NP
  problems},}\ \href {\doibase 10.3389/fphy.2014.00005} {\bibfield  {journal}
  {\bibinfo  {journal} {Front. Phys.}\ }\textbf {\bibinfo {volume} {2}},\
  \bibinfo {pages} {5} (\bibinfo {year} {2014})}\BibitemShut {NoStop}%
\bibitem [{\citenamefont {Albash}\ and\ \citenamefont
  {Lidar}(2018)}]{albash2018adiabatic}%
  \BibitemOpen
  \bibfield  {author} {\bibinfo {author} {\bibfnamefont {T.}~\bibnamefont
  {Albash}}\ and\ \bibinfo {author} {\bibfnamefont {D.~A.}\ \bibnamefont
  {Lidar}},\ }\enquote {\bibinfo {title} {Adiabatic quantum computation},}\
  \href {\doibase 10.1103/RevModPhys.90.015002} {\bibfield  {journal} {\bibinfo
   {journal} {Rev. Mod. Phys.}\ }\textbf {\bibinfo {volume} {90}},\ \bibinfo
  {pages} {015002} (\bibinfo {year} {2018})}\BibitemShut {NoStop}%
\bibitem [{\citenamefont {Kolodrubetz}\ \emph {et~al.}(2017)\citenamefont
  {Kolodrubetz}, \citenamefont {Sels}, \citenamefont {Mehta},\ and\
  \citenamefont {Polkovnikov}}]{kolodrubetz2017geometry}%
  \BibitemOpen
  \bibfield  {author} {\bibinfo {author} {\bibfnamefont {M.}~\bibnamefont
  {Kolodrubetz}}, \bibinfo {author} {\bibfnamefont {D.}~\bibnamefont {Sels}},
  \bibinfo {author} {\bibfnamefont {P.}~\bibnamefont {Mehta}}, \ and\ \bibinfo
  {author} {\bibfnamefont {A.}~\bibnamefont {Polkovnikov}},\ }\enquote
  {\bibinfo {title} {Geometry and non-adiabatic response in quantum and
  classical systems},}\ \href {\doibase 10.1016/j.physrep.2017.07.001}
  {\bibfield  {journal} {\bibinfo  {journal} {Phys. Rep.}\ }\textbf {\bibinfo
  {volume} {697}},\ \bibinfo {pages} {1} (\bibinfo {year} {2017})}\BibitemShut
  {NoStop}%
\bibitem [{\citenamefont {del Campo}\ \emph {et~al.}(2012)\citenamefont {del
  Campo}, \citenamefont {Rams},\ and\ \citenamefont
  {Zurek}}]{delcampo2012assisted}%
  \BibitemOpen
  \bibfield  {author} {\bibinfo {author} {\bibfnamefont {A.}~\bibnamefont {del
  Campo}}, \bibinfo {author} {\bibfnamefont {M.~M.}\ \bibnamefont {Rams}}, \
  and\ \bibinfo {author} {\bibfnamefont {W.~H.}\ \bibnamefont {Zurek}},\
  }\enquote {\bibinfo {title} {Assisted Finite-Rate Adiabatic Passage Across a
  Quantum Critical Point: Exact Solution for the Quantum Ising Model},}\ \href
  {\doibase 10.1103/PhysRevLett.109.115703} {\bibfield  {journal} {\bibinfo
  {journal} {Phys. Rev. Lett.}\ }\textbf {\bibinfo {volume} {109}},\ \bibinfo
  {pages} {115703} (\bibinfo {year} {2012})}\BibitemShut {NoStop}%
\bibitem [{\citenamefont {Alipour}\ \emph {et~al.}(2020)\citenamefont
  {Alipour}, \citenamefont {Chenu}, \citenamefont {Rezakhani},\ and\
  \citenamefont {del Campo}}]{Alipour2020shortcutsto}%
  \BibitemOpen
  \bibfield  {author} {\bibinfo {author} {\bibfnamefont {S.}~\bibnamefont
  {Alipour}}, \bibinfo {author} {\bibfnamefont {A.}~\bibnamefont {Chenu}},
  \bibinfo {author} {\bibfnamefont {A.~T.}\ \bibnamefont {Rezakhani}}, \ and\
  \bibinfo {author} {\bibfnamefont {A.}~\bibnamefont {del Campo}},\ }\enquote
  {\bibinfo {title} {Shortcuts to {A}diabaticity in {D}riven {O}pen {Q}uantum
  {S}ystems: {B}alanced {G}ain and {L}oss and {N}on-{M}arkovian {E}volution},}\
  \href {\doibase 10.22331/q-2020-09-28-336} {\bibfield  {journal} {\bibinfo
  {journal} {{Quantum}}\ }\textbf {\bibinfo {volume} {4}},\ \bibinfo {pages}
  {336} (\bibinfo {year} {2020})}\BibitemShut {NoStop}%
\bibitem [{\citenamefont {Dann}\ \emph {et~al.}(2020)\citenamefont {Dann},
  \citenamefont {Tobalina},\ and\ \citenamefont {Kosloff}}]{dann2020fast}%
  \BibitemOpen
  \bibfield  {author} {\bibinfo {author} {\bibfnamefont {R.}~\bibnamefont
  {Dann}}, \bibinfo {author} {\bibfnamefont {A.}~\bibnamefont {Tobalina}}, \
  and\ \bibinfo {author} {\bibfnamefont {R.}~\bibnamefont {Kosloff}},\
  }\enquote {\bibinfo {title} {Fast route to equilibration},}\ \href {\doibase
  10.1103/PhysRevA.101.052102} {\bibfield  {journal} {\bibinfo  {journal}
  {Phys. Rev. A}\ }\textbf {\bibinfo {volume} {101}},\ \bibinfo {pages}
  {052102} (\bibinfo {year} {2020})}\BibitemShut {NoStop}%
\bibitem [{\citenamefont {Dann}\ \emph {et~al.}(2019)\citenamefont {Dann},
  \citenamefont {Tobalina},\ and\ \citenamefont {Kosloff}}]{dann2019shortcut}%
  \BibitemOpen
  \bibfield  {author} {\bibinfo {author} {\bibfnamefont {R.}~\bibnamefont
  {Dann}}, \bibinfo {author} {\bibfnamefont {A.}~\bibnamefont {Tobalina}}, \
  and\ \bibinfo {author} {\bibfnamefont {R.}~\bibnamefont {Kosloff}},\
  }\enquote {\bibinfo {title} {Shortcut to Equilibration of an Open Quantum
  System},}\ \href {\doibase 10.1103/PhysRevLett.122.250402} {\bibfield
  {journal} {\bibinfo  {journal} {Phys. Rev. Lett.}\ }\textbf {\bibinfo
  {volume} {122}},\ \bibinfo {pages} {250402} (\bibinfo {year}
  {2019})}\BibitemShut {NoStop}%
\bibitem [{\citenamefont {Das}\ and\ \citenamefont
  {Mukherjee}(2020)}]{das2020quantum}%
  \BibitemOpen
  \bibfield  {author} {\bibinfo {author} {\bibfnamefont {A.}~\bibnamefont
  {Das}}\ and\ \bibinfo {author} {\bibfnamefont {V.}~\bibnamefont
  {Mukherjee}},\ }\enquote {\bibinfo {title} {Quantum-enhanced finite-time Otto
  cycle},}\ \href {\doibase 10.1103/PhysRevResearch.2.033083} {\bibfield
  {journal} {\bibinfo  {journal} {Phys. Rev. Research}\ }\textbf {\bibinfo
  {volume} {2}},\ \bibinfo {pages} {033083} (\bibinfo {year}
  {2020})}\BibitemShut {NoStop}%
\bibitem [{\citenamefont {Johansson}\ \emph {et~al.}(2013)\citenamefont
  {Johansson}, \citenamefont {Nation},\ and\ \citenamefont
  {Nori}}]{johannson2013qutip}%
  \BibitemOpen
  \bibfield  {author} {\bibinfo {author} {\bibfnamefont {J.~R.}\ \bibnamefont
  {Johansson}}, \bibinfo {author} {\bibfnamefont {P.~D.}\ \bibnamefont
  {Nation}}, \ and\ \bibinfo {author} {\bibfnamefont {F.}~\bibnamefont
  {Nori}},\ }\enquote {\bibinfo {title} {QuTiP 2: A Python framework for the
  dynamics of open quantum systems},}\ \href {\doibase
  10.1016/j.cpc.2012.11.019} {\bibfield  {journal} {\bibinfo  {journal}
  {Comput. Phys. Commun.}\ }\textbf {\bibinfo {volume} {184}},\ \bibinfo
  {pages} {1234 } (\bibinfo {year} {2013})}\BibitemShut {NoStop}%
\bibitem [{\citenamefont {Abah}\ and\ \citenamefont
  {Lutz}(2017)}]{abah2017energy}%
  \BibitemOpen
  \bibfield  {author} {\bibinfo {author} {\bibfnamefont {O.}~\bibnamefont
  {Abah}}\ and\ \bibinfo {author} {\bibfnamefont {E.}~\bibnamefont {Lutz}},\
  }\enquote {\bibinfo {title} {Energy efficient quantum machines},}\ \href
  {\doibase 10.1209/0295-5075/118/40005} {\bibfield  {journal} {\bibinfo
  {journal} {{EPL}}\ }\textbf {\bibinfo {volume} {118}},\
  \bibinfo {pages} {40005} (\bibinfo {year} {2017})}\BibitemShut {NoStop}%
\bibitem [{\citenamefont {Campbell}\ and\ \citenamefont
  {Deffner}(2017)}]{campbell2017tradeoff}%
  \BibitemOpen
  \bibfield  {author} {\bibinfo {author} {\bibfnamefont {S.}~\bibnamefont
  {Campbell}}\ and\ \bibinfo {author} {\bibfnamefont {S.}~\bibnamefont
  {Deffner}},\ }\enquote {\bibinfo {title} {Trade-Off Between Speed and Cost in
  Shortcuts to Adiabaticity},}\ \href {\doibase 10.1103/PhysRevLett.118.100601}
  {\bibfield  {journal} {\bibinfo  {journal} {Phys. Rev. Lett.}\ }\textbf
  {\bibinfo {volume} {118}},\ \bibinfo {pages} {100601} (\bibinfo {year}
  {2017})}\BibitemShut {NoStop}%
\bibitem [{\citenamefont {Zheng}\ \emph {et~al.}(2016)\citenamefont {Zheng},
  \citenamefont {Campbell}, \citenamefont {De~Chiara},\ and\ \citenamefont
  {Poletti}}]{zheng2016cost}%
  \BibitemOpen
  \bibfield  {author} {\bibinfo {author} {\bibfnamefont {Y.}~\bibnamefont
  {Zheng}}, \bibinfo {author} {\bibfnamefont {S.}~\bibnamefont {Campbell}},
  \bibinfo {author} {\bibfnamefont {G.}~\bibnamefont {De~Chiara}}, \ and\
  \bibinfo {author} {\bibfnamefont {D.}~\bibnamefont {Poletti}},\ }\enquote
  {\bibinfo {title} {Cost of counterdiabatic driving and work output},}\ \href
  {\doibase 10.1103/PhysRevA.94.042132} {\bibfield  {journal} {\bibinfo
  {journal} {Phys. Rev. A}\ }\textbf {\bibinfo {volume} {94}},\ \bibinfo
  {pages} {042132} (\bibinfo {year} {2016})}\BibitemShut {NoStop}%
\bibitem [{\citenamefont {Tobalina}\ \emph {et~al.}(2019)\citenamefont
  {Tobalina}, \citenamefont {Lizuain},\ and\ \citenamefont
  {Muga}}]{tobalina2019vanishing}%
  \BibitemOpen
  \bibfield  {author} {\bibinfo {author} {\bibfnamefont {A.}~\bibnamefont
  {Tobalina}}, \bibinfo {author} {\bibfnamefont {I.}~\bibnamefont {Lizuain}}, \
  and\ \bibinfo {author} {\bibfnamefont {J.~G.}\ \bibnamefont {Muga}},\
  }\enquote {\bibinfo {title} {Vanishing efficiency of a speeded-up
  ion-in-Paul-trap Otto engine},}\ \href {\doibase 10.1209/0295-5075/127/20005}
  {\bibfield  {journal} {\bibinfo  {journal} {{EPL}}\
  }\textbf {\bibinfo {volume} {127}},\ \bibinfo {pages} {20005} (\bibinfo
  {year} {2019})}\BibitemShut {NoStop}%
\bibitem [{\citenamefont {Manatuly}\ \emph {et~al.}(2019)\citenamefont
  {Manatuly}, \citenamefont {Niedenzu}, \citenamefont {Rom\'an-Ancheyta},
  \citenamefont {\c{C}akmak}, \citenamefont {M\"ustecapl\ifmmode \imath \else
  \i \fi{}o\ifmmode~\breve{g}\else \u{g}\fi{}lu},\ and\ \citenamefont
  {Kurizki}}]{manatuly2019collectively}%
  \BibitemOpen
  \bibfield  {author} {\bibinfo {author} {\bibfnamefont {A.}~\bibnamefont
  {Manatuly}}, \bibinfo {author} {\bibfnamefont {W.}~\bibnamefont {Niedenzu}},
  \bibinfo {author} {\bibfnamefont {R.}~\bibnamefont {Rom\'an-Ancheyta}},
  \bibinfo {author} {\bibfnamefont {B.}~\bibnamefont {\c{C}akmak}}, \bibinfo
  {author} {\bibfnamefont {{\"O}.~E.}\ \bibnamefont {M\"ustecapl\ifmmode \imath
  \else \i \fi{}o\ifmmode~\breve{g}\else \u{g}\fi{}lu}}, \ and\ \bibinfo
  {author} {\bibfnamefont {G.}~\bibnamefont {Kurizki}},\ }\enquote {\bibinfo
  {title} {Collectively enhanced thermalization via multiqubit collisions},}\
  \href {\doibase 10.1103/PhysRevE.99.042145} {\bibfield  {journal} {\bibinfo
  {journal} {Phys. Rev. E}\ }\textbf {\bibinfo {volume} {99}},\ \bibinfo
  {pages} {042145} (\bibinfo {year} {2019})}\BibitemShut {NoStop}%
\bibitem [{\citenamefont {Funo}\ \emph {et~al.}(2017)\citenamefont {Funo},
  \citenamefont {Zhang}, \citenamefont {Chatou}, \citenamefont {Kim},
  \citenamefont {Ueda},\ and\ \citenamefont {del Campo}}]{funo2017universal}%
  \BibitemOpen
  \bibfield  {author} {\bibinfo {author} {\bibfnamefont {K.}~\bibnamefont
  {Funo}}, \bibinfo {author} {\bibfnamefont {J.-N.}\ \bibnamefont {Zhang}},
  \bibinfo {author} {\bibfnamefont {C.}~\bibnamefont {Chatou}}, \bibinfo
  {author} {\bibfnamefont {K.}~\bibnamefont {Kim}}, \bibinfo {author}
  {\bibfnamefont {M.}~\bibnamefont {Ueda}}, \ and\ \bibinfo {author}
  {\bibfnamefont {A.}~\bibnamefont {del Campo}},\ }\enquote {\bibinfo {title}
  {Universal Work Fluctuations During Shortcuts to Adiabaticity by
  Counterdiabatic Driving},}\ \href {\doibase 10.1103/PhysRevLett.118.100602}
  {\bibfield  {journal} {\bibinfo  {journal} {Phys. Rev. Lett.}\ }\textbf
  {\bibinfo {volume} {118}},\ \bibinfo {pages} {100602} (\bibinfo {year}
  {2017})}\BibitemShut {NoStop}%
\end{thebibliography}
\end{document}